%% file: DealMaTe_main.tex
\documentclass[acmtog]{acmart}

\renewcommand\footnotetextcopyrightpermission[1]{} 

\usepackage{xspace}
\usepackage{booktabs} 
\usepackage{enumitem}
\usepackage{makecell} 

\usepackage{multirow}

\AtBeginDocument{%
  }

\usepackage{booktabs}
\usepackage{caption}
\usepackage{subcaption}
\usepackage{multirow}
\usepackage{enumitem}
\usepackage{fmtcount}
\usepackage{microtype}
\usepackage{balance}

\usepackage[ruled]{algorithm2e} 
\usepackage{makecell}
\usepackage{xcolor} 
\usepackage{array}
\usepackage[table]{xcolor}
\definecolor{LightBlue}{rgb}{0.9,0.94,1}
\definecolor{blue}{RGB}{0, 0, 255}
\definecolor{red}{RGB}{255, 0, 0}

\definecolor{marron}{HTML}{CC88B0}
\acmISBN{978-1-4503-XXXX-X/2018/06}

\makeatletter
\def\@fnsymbol#1{\ensuremath{\ifcase#1\or \dagger\or \ddagger\or
   \mathsection\or \mathparagraph\or \|\or **\or \dagger\dagger
   \or \ddagger\ddagger \else\@ctrerr\fi}}
\makeatother

\begin{document}

\title{\textcolor{black}{DealMaTe: Multi-Dimensional Material Transfer via Diffusion Transformer}}

\author{Nisha Huang}
\orcid{0000-0002-1627-6584}
 \affiliation{
 \institution{Tsinghua University}
 \country{China}
 }
 \affiliation{
 \institution{Pengcheng Laboratory}
 \country{China}
 }
\email{hns24@mails.tsinghua.edu.cn}

 \author{Yizhou Lin}
 \orcid{0009-0001-7716-7437}
 \affiliation{
 \institution{Tsinghua University}
  \country{China}
 }
 \email{yz-lin24@mails.tsinghua.edu.cn}

\author{Jie Guo}
\orcid{0000-0002-7411-4751}
\affiliation{
\institution{Pengcheng Laboratory}
\country{China}
}
\email{guoj01@pcl.ac.cn}

\author{Xiu Li}
\orcid{0000-0003-0403-1923}
\affiliation{%
  \institution{Tsinghua University}
  \country{China}
}
\authornote{Co-corresponding authors} 
\email{li.xiu@sz.tsinghua.edu.cn}

\author{Tong-Yee Lee}
\orcid{0000-0001-6699-2944}
\affiliation{
 \institution{National Cheng-Kung University}
 \country{Taiwan}
}
\authornotemark[1] 
\email{tonylee@ncku.edu.tw}

\author{Zitong Yu}
\orcid{0000-0003-0422-6616}
\affiliation{%
  \institution{Great Bay University}
  \country{China}
}
 \affiliation{
 \institution{Dongguan Key Laboratory for Intelligence and Information Technology}
 \country{China}
 }
\authornotemark[1] 
\email{zitong.yu@ieee.org}

\renewcommand{\shortauthors}{Huang et al.}


\input{Sections/0_abstract}



\begin{teaserfigure}
  \centering
   \includegraphics[width=\linewidth]{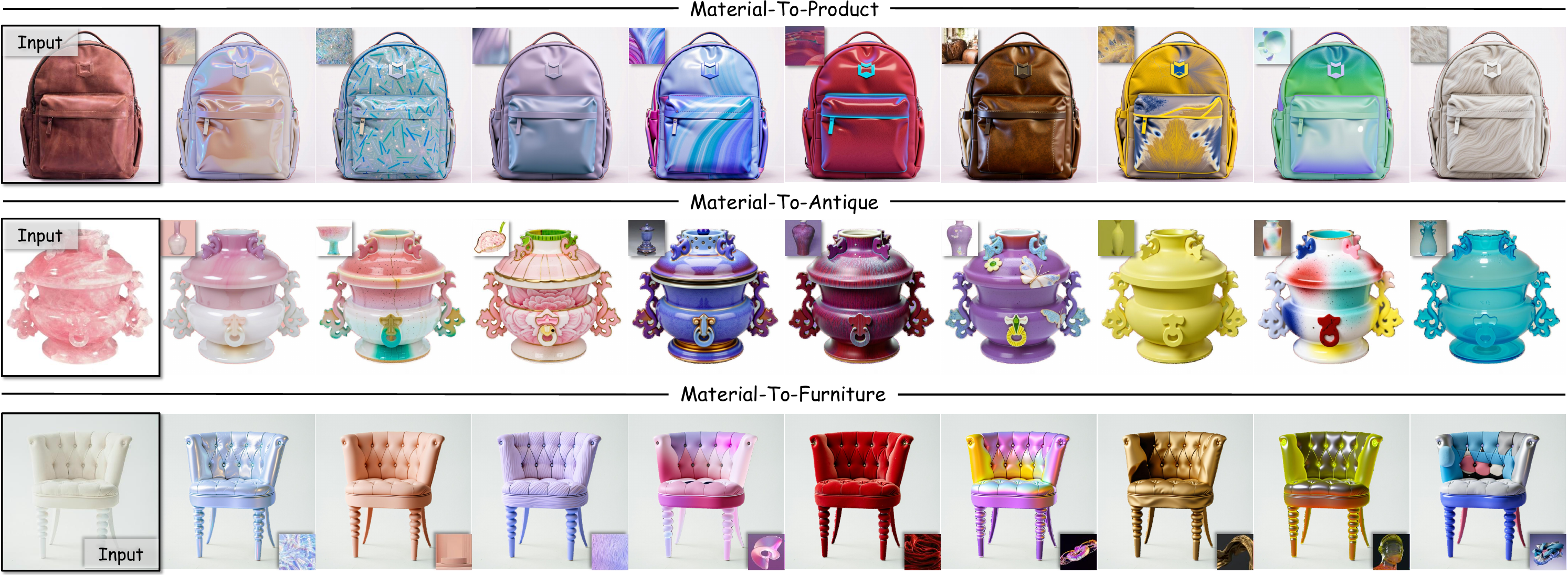}
   \caption{\textbf{DealMaTe} is a material transfer method that can transform materials from a single real-world image without any prior knowledge. This method is not only capable of successfully extracting texture information from antiques with thousands of years of history but also handles novel virtual materials generated by computer graphics images. It excels across diverse scenarios such as product design, antique restoration, and furniture design, thereby robustly facilitating design tasks.}
  \label{fig:teaser}
\end{teaserfigure}

\maketitle

\input{Sections/1_intro}

\input{Sections/2_related}

\input{Sections/3_method}

\input{Sections/4_experiments}

\input{Sections/5_Discussion}
\input{Sections/6_conclusion}

\begin{acks}
This work is supported by the Science and Technology Innovation 2030-Major Projects (Grant No. 2021ZD0201404), the Shenzhen Key Laboratory of New Generation Interactive Media Technology Innovation (Grant No. ZDSYS20210623092001004), 
the National Science and Technology
Council, Taiwan (Grant No. 114-2221-E-006-114-MY3),
the National Natural Science Foundation of China (Grant Nos. 62306061 and 62576076), and is sponsored by the CCF-Tencent Rhino-Bird Open Research Fund. 
Nisha Huang is supported by the Doctoral Student Program of the Young Science and Technology Talents Cultivation Project, China Association for Science and Technology.
\end{acks}

\bibliographystyle{ACM-Reference-Format}
\bibliography{sample-base}

\end{document}

%% file: Sections/0_abstract.tex
\begin{abstract}
\input{Figures/insight}
Recently, diffusion-based material transfer methods rely on image fine-tuning or complex architectures with auxiliary networks but face challenges such as text dependency, additional computational costs, and feature misalignment. 
To address these limitations, we propose \textbf{DealMaTe}, using \underline{\textbf{de}}pth, norm\underline{\textbf{a}}l, and \underline{\textbf{l}}ighting images for \underline{\textbf{ma}}terial \underline{\textbf{t}}ransf\underline{\textbf{e}}r.
DealMaTe is a simplified diffusion framework that eliminates text guidance and reference networks.
We design a lightweight 3D information injection method, Multi-Dim 3D Shader LoRA, which, without modifying the base model weights, enables compatible control conditions and achieves harmonious and stable results. Additionally, we optimize the attention mechanism with Shader Causal Mutual Attention and key-value (KV) caching to reduce inference latency caused by multiple conditions, improve computational efficiency, and achieve high-quality material transfer results with low architectural complexity. Extensive experiments covering a wide variety of objects and lighting conditions consistently demonstrate that DealMaTe achieves remarkable high-fidelity material transfer under arbitrary input materials.
The code is available at \url{https://github.com/haha-lisa/DealMaTe}.
\end{abstract}

%% file: Figures/insight.tex
\begin{figure*}[!h]
\centering
\includegraphics[width= 0.9\linewidth]{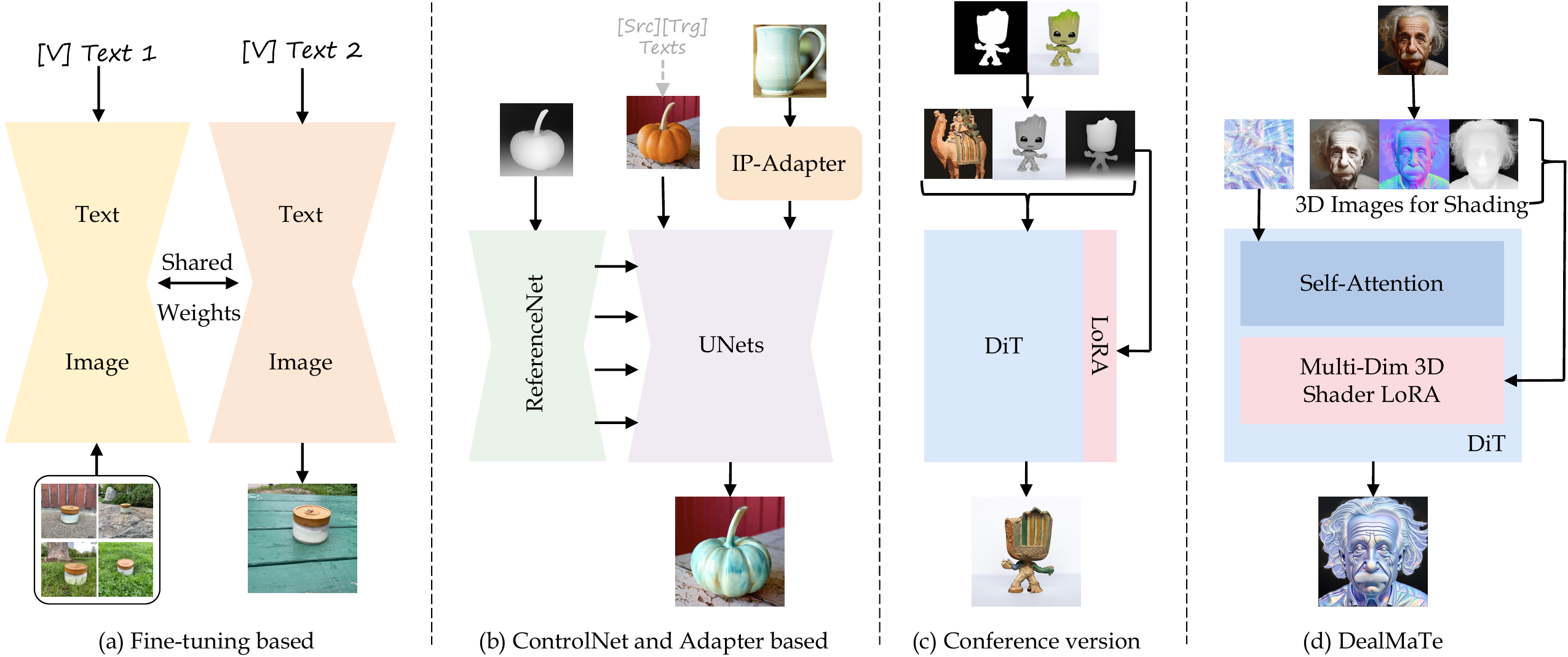}
\caption{Simplified structure comparison of different kinds of material transfer methods. Our approach neither relies on fine-tuning image sets/individual images nor requires additional image encoding by IP-Adapters. It only requires basic image information as input, rather than complex text guidance, to obtain high-quality material transfer results.}
\label{fig:insight}
\end{figure*}

%% file: Sections/1_intro.tex
\section{Introduction}
\label{sec:intro}

Material transfer is a technique that precisely maps the properties of a specific material sample onto the surface of a target object, as shown in Fig.~\ref{fig:teaser}. 
Due to its broad application prospects in digital content creation and industrial design, this technology has garnered attention in recent years.
Traditional material transfer frameworks primarily rely on parametric modeling paradigms, such as optical reflection models based on bidirectional scattering distribution function (BRDF)~\cite{deschaintre2018single,hu2022inverse,henzler2021generative} or procedural texture generation algorithms~\cite{tony,hu2019novel}. These frameworks limit the diversity of the generated results due to the limited size of the basic material library, failing to meet the customization needed for non-uniform composite materials in digital artistic creation.

In recent years, inspired by the breakthroughs of diffusion models in conditional generation tasks, diffusion-based material transfer methods~\cite{sharma2024alchemist, wu2024u, lopes2024material} have achieved remarkable improvements in high-quality material transfer effects. Current state-of-the-art (SOTA) approaches often employ fine-tuning of diffusion models with sample sets bound to text identifiers~\cite{ruiz2023dreambooth,zhang2023prospect,wu2024u} (in Fig.~\ref{fig:insight} (a)), implicitly encoding material features through conceptual semantic binding. Although these methods address some shortcomings of traditional pipelines, their heavy reliance on text prompts restricts fine-grained control of material properties. Moreover, the full-parameter fine-tuning paradigm significantly increases training costs and the risk of overfitting. 
The latest studies~\cite{garifullin2025materialfusion, cheng2024zest,cheng2025marble} have also introduced pre-trained general image encoders IP-Adapters~\cite{ye2023ip-adapter} to extract material features, as well as ControlNet~\cite{zhang2023adding} to inject depth information, as shown in Fig.~\ref{fig:insight} (b). 
However, these methods frequently induce hierarchical decoupling between material and structural information (instead of seamless fusion) during generation, while also suffering from prolonged inference times.
\textcolor{black}{
Fig.~\ref{fig:insight} (c) eliminates the use of additional image encoders in material transfer tasks. MaTe prioritizes the semantic alignment of cross-modal features and their interaction in a shared latent space, mapping materials, depth, and grayscale images to a unified latent representation. However, our previous work~\cite{Huang_2025_ICCV} and other SOTAs~\cite{cheng2024zest,cheng2025marble,garifullin2025materialfusion} only employ Depth LoRA for structural control and use grayscale images as initial sampling images for lighting control, resulting in insufficient preservation of 3D information.
}

Therefore, to solve the problem, DealMaTe (in Fig.~\ref{fig:insight} (d)) designs the Multi-Dim 3D Shader LoRA: depth, normal, and lighting. Specifically, the depth LoRA is capable of precisely capturing the depth information of the target object, providing an accurate three-dimensional spatial reference for material transfer. The normal LoRA focuses on controlling surface curvature, ensuring that material textures closely conform to the geometric shape of the target object, thereby effectively avoiding the general problem~\cite{cheng2024zest,cheng2025marble,garifullin2025materialfusion} of material-geometry misalignment. 
To the best of our knowledge, we pioneer the fine-tuning of the Lighting LoRA in the image generation field.
It finely adjusts the direction, intensity, and color of the lighting, enabling specular highlights to accurately follow the direction of the light source and significantly improving the consistency of lighting and reflection in the material transfer results.

We suggest a lightweight conditional injection LoRA module at the model architecture level to address the problem of multi-condition collaborative control.
Our essential innovation is the solitary insertion of conditional signals. 
By means of a parallel-branch design, the module is grafted onto the frozen backbone so that low-rank projection is restricted to the conditional-branch tokens, whereas the prompt and noise pathways stay locked and unchanged.
This advanced architecture facilitates that the Multi-Dim 3D Shader LoRA works seamlessly with the customized model while also allowing for harmonic and strong zero-shot generalization.

Regarding the inference latency caused by multiple conditions, at the attention mechanism level, we replace the traditional full attention with our innovative Shader Causal Mutual Attention and combine it with KV caching technology to achieve the transformation.
At the initial diffusion step, the system computes the key–value tensors for every conditioning feature once and stores them. These cached tensors are then reused in all subsequent steps, cutting out repeated computation and delivering significant savings.


In summary, we highlight our contributions as follows:

\begin{itemize}
    \item We propose DealMaTe, an efficient and concise material transfer architecture that can infer all necessary information using only image conditions, without the need for fine-tuning or manual text settings, and supports zero-shot generalization.
    \item We introduce Multi-Dim 3D Shader LoRA to address material-geometry misalignment and lighting-reflection inconsistency. To the best of our knowledge, we are the first to fine-tune the Lighting LoRA. We also design a lightweight conditional injection LoRA module that isolates the injection of conditional signals, integrates seamlessly with pre-trained models, and enhances the ability for multi-condition collaborative control.
    \item We optimize the attention mechanism through Shader Causal Mutual Attention and KV caching to reduce inference latency caused by various conditions, improve computational efficiency, and reduce computational resources, achieving high-quality material transfer results with lower architectural complexity.
\end{itemize}

This work extends our recent conference paper MaTe published in ICCV 2025~\cite{Huang_2025_ICCV}. 
The material transfer framework has been enhanced by adjusting the injection method of Low-Rank Adaptation (LoRA)~\cite{hulora} and proposing Multi-Dim 3D Shader LoRA, which introduces brand-new control conditions for surface curvature and lighting. Furthermore, to improve the harmonious stability of multi-condition collaboration, we have adopted Shader Causal Mutual Attention and KV caching to optimize the attention mechanism. The extended version has been thoroughly evaluated through comprehensive experiments, which show that our DealMaTe can greatly improve the quality of findings for existing arbitrary image material transfer models. 
For visual comparisons between MaTe and DealMaTe, please refer to Fig.~\ref{fig:insight-comparison}.

%% file: Sections/2_related.tex
\input{Figures/insight-comparison}
\input{Figures/pipeline}

\section{Related Work}
\paragraph{Material and texture extraction.}
Material acquisition has always been an ongoing challenge~\cite{Guarnera16}. Traditional material capture often relies on costly multi-view~\cite{asselin2020deep} or polarized light~\cite{deschaintre2021deep} equipment. Many researchers have used synthetic datasets to train networks that estimate SVBRDF from a single viewpoint~\cite{deschaintre2018single}, often in conjunction with additional single-view data~\cite{gao2019deep,martin2022materia} or specific training strategies~\cite{li2017modeling,vecchio2021surfacenet,Deschaintre2020guided}. 
However, it is important to note that all the aforementioned studies require a frontal view of the material, meaning the camera must be parallel to the material surface, which is often difficult to achieve in practical applications. 
For material analysis, UMat~\cite{rodriguez2023umat} employs a single image taken using a flatbed scanner.
Rather than depicting the substance itself, TexSynth~\cite{texsynth} offers a guided texture modification method.
Material Palette~\cite{lopes2024material} suggests a way to extract materials from a single image without previous knowledge.
MaterialPicker~\cite{ma2025materialpicker} uses a DiT-based text-to-video generation methodology to generate material maps.
Moreover, a common way to extract textures from real-world images includes clustering the textures and scaling them up to full resolution~\cite{rosenberger2009layered,li2022scraping} or applying dataset distillation techniques~\cite{cazenavette2022wearable}.

\paragraph{Image-guided generation.}
Image-guided generation is a challenging yet promising field, leveraging visual guidance for content creation. 
The emergence of diffusion models~\cite{rombach2022high,peebles2023scalable,huang2024diffstyler} has significantly propelled advancements in this field, enabling SOTA performance across a wide range of generative visual tasks, such as image-to-image translation~\cite{ye2023ip-adapter,huang2025creativesynth}, subject-driven image generation~\cite{ruiz2023dreambooth,hertz2022prompt}, etc.
DreamBooth~\cite{ruiz2023dreambooth} and Textual Inversion~\cite{galimage} use transfer learning in text-to-image (T2I) diffusion models to generate customized concepts through parameter fine-tuning or word vector optimization.
To enhance the controllability of image-guided generation, adapter-based architectures have emerged as bridges between external control signals (e.g., sketches) and diffusion models.
ControlNet~\cite{zhang2023adding} validates an adapter that can be trained to capture task-specific input conditions, whereas T2I-adapter~\cite{mou2024t2i} uses a lightweight adapter to achieve fine-grained control in the color and structure of the produced images.
IP-Adapter~\cite{ye2023ip-adapter} allows for more flexible and intuitive control of the generation process, expanding the capabilities of image-guided generation. 
These methods demonstrate image-conditioned generation strategies with varying specificity, suggesting potential pathways for addressing material transfer challenges.

\paragraph{Material acquisition and transfer.}
Material acquisition and transfer represent a realm of research considering illumination conditions, object geometry, and physical properties of materials. 
Traditional 3D material transfer methods, such as Text2tex~\cite{chen2023text2tex}, TEXTure~\cite{richardson2023texture}, and TextureDreamer~\cite{yeh2024texturedreamer}, rely on 3D geometric shapes and lighting estimation, followed by careful adjustment of material properties.
As a result, the quality and diversity are restricted, presenting limited and unsatisfactory results.
In contrast, 2D-to-2D material transfer, bypassing ground-truth 3D data, is challenging yet highly practical.
Prevalent approaches based on fine-tuning, including DreamBooth~\cite{ruiz2023dreambooth}, Material Palette~\cite{lopes2024material}, Prospect~\cite{zhang2023prospect}, and U-VAP~\cite{wu2024u} fine-tune diffusion models on small sample sets associated with text identifiers. 
While existing methods mitigate conventional pipeline limitations, text prompts reliance restricts control and fine-tuning risks computational costs, and overfitting. 
Another category of methods, such as MaterialFusion~\cite{garifullin2025materialfusion} and ZeST~\cite{cheng2024zest}, uses extra encoder modules like IP-Adapter~\cite{ye2023ip-adapter} or ControlNet~\cite{zhang2023adding} to extract material features.
When multiple image conditions are injected into the network in parallel, the material and structure information in the results are separated into two independent layers, lacking semantic integration (see Fig.~\ref{fig:comparison} (d)-(g)). The latest SOTA method MaterialFusion~\cite{garifullin2025materialfusion} applies DDIM inversion~\cite{songdenoising} to both the material and the input images, which increases the sampling steps and time by more than double.
Different from methods that rely on text-assisted fine-tuning or complex architectures with additional reference networks, our method DealMaTe projects material, content, and depth images into the same latent space and performs semantically aligned generation.

%% file: Figures/insight-comparison.tex
\begin{figure}
\centering
\includegraphics[width= \linewidth]{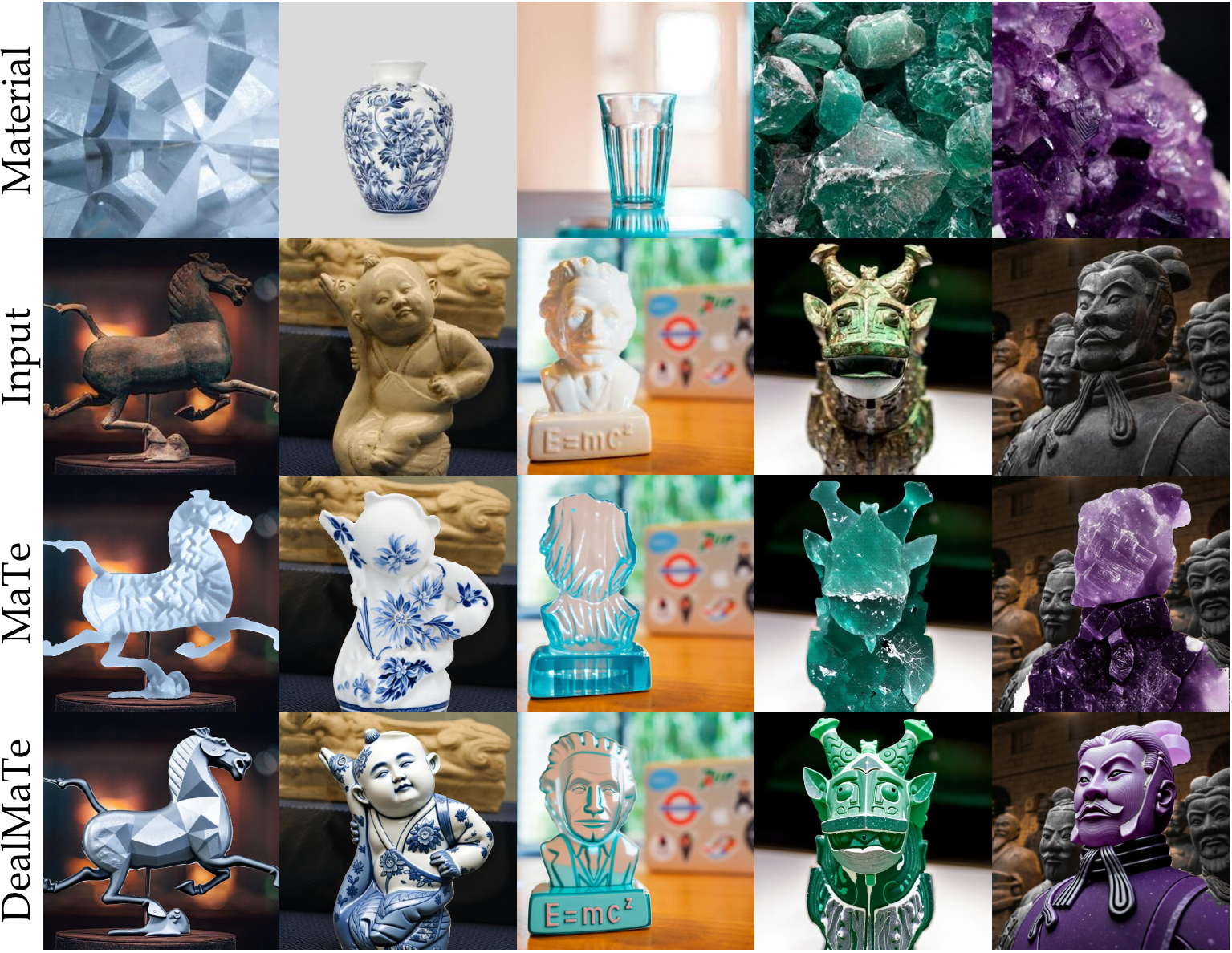}
\caption{Comparison with our conference version work MaTe. DealMaTe is capable of faithfully transferring material characteristics while ensuring enhanced structural consistency with the input image.}
\label{fig:insight-comparison}
\end{figure}

%% file: Figures/pipeline.tex
\begin{figure*}
\centering
\includegraphics[width=\linewidth]{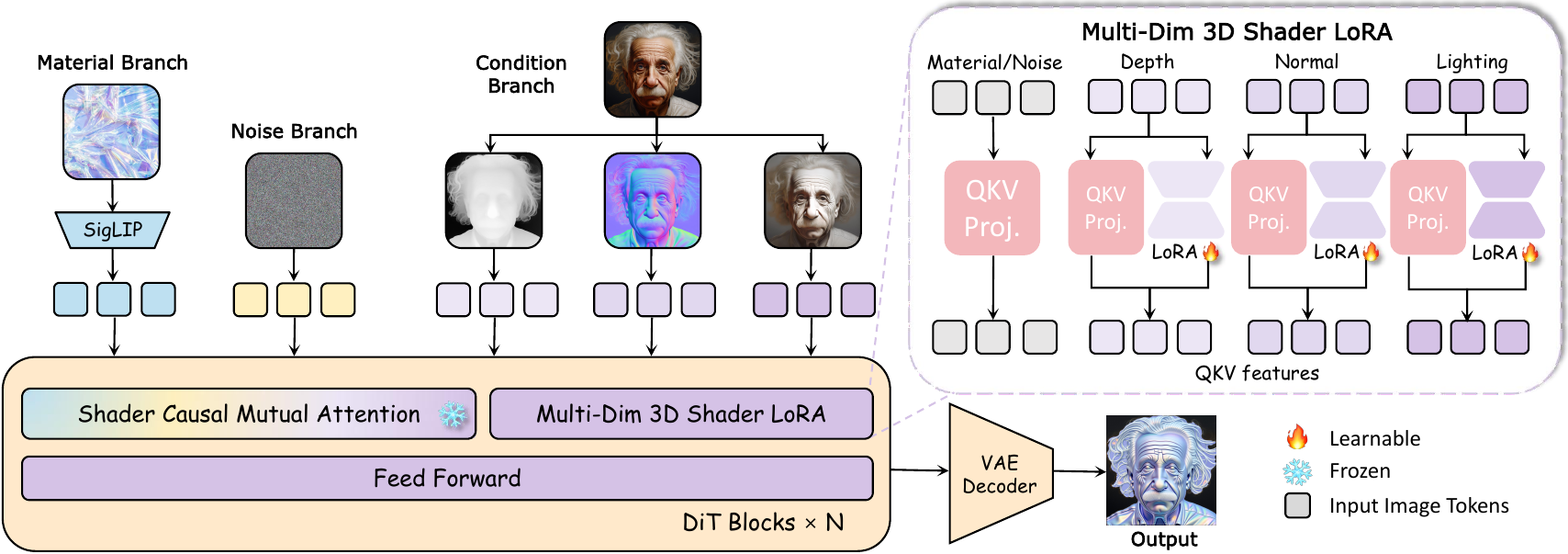}
\caption{Our method achieves high-quality material transfer by feeding depth, normal, and lighting inputs into respective shader LoRAs and then passing the material and 3D condition tokens into the shader causal mutual attention for interaction, ensuring that they remain in the same feature space throughout the diffusion process. It remarkably eliminates the need for unnecessary image fine-tuning and text guidance, resulting in a streamlined inference process.
}

\label{fig:pipeline}
\end{figure*}

%% file: Sections/3_method.tex
\section{Method}
Building on the Diffusion Transformer architecture introduced in Section~\ref{sec:3.1}, we maintain spatial consistency through rotary position encoding and integrate multiple conditional signals via the multimodal attention mechanism. For conditional feature enhancement, the Multi-Dim 3D Shader LoRA proposed in Section~\ref{sec:3.2} selectively boosts conditional representations while preserving the base model's material representation capabilities. During inference, the Shader Causal Mutual Attention mechanism designed in Section~\ref{sec:3.3} efficiently fuses multiple conditions and effectively prevents cross-condition interference. The training process employs the flow-matching loss function detailed in Section~\ref{sec:3.4}, and inference efficiency is optimized through Key-Value (KV) caching techniques that eliminate redundant computations across denoising steps. 
Additionally, the data collection process outlined in Section~\ref{sec:3.5} leverages Marigold-based pipelines to obtain high-quality depth, normal, and lighting estimates from natural images.
The simultaneous estimation of these three intrinsic properties requires 0.3 seconds for per input image.

\subsection{ Preliminary }
\label{sec:3.1}

\subsubsection{Rectified-Flow Models}
Generative models seek to establish a correspondence between samples \( x_1 \) originating from a noise distribution \( p_1 \) and instances \( x_0 \) derived from a target distribution \( p_0 \), where \( p_0 \) corresponds to real images.
According to~\cite{lipmanflow, liuflow}, rectified flows establish a forward procedure that constructs linear pathways across distributions \(p_0\) and \(p_1\), with \(p_1\) defined as \(\mathcal{N}(0, 1)\) in Eq.~\ref{eqn:flow_path}. This method is time-reliant given the presence of the timestep \(t\).
\begin{equation}
    \label{eqn:flow_path}
    x_t = (1 - t)x_0 + t\epsilon, \quad \epsilon \sim N(0, 1).
\end{equation}
The acquisition of this transformation is achieved by optimizing a neural model with parameters $\theta$ to estimate the rectified flow velocity \(v\), denoted \(v_\theta\). Through the parameterization approach from~\cite{esser2024scaling}, this network that estimates flow velocity can alternatively operate as a model for predicting noise, \(\epsilon_\theta\), which is trained with the Conditional Flow Matching (CFM) objective detailed in Eq.~\ref{eqn:cfm_objective}.
\begin{equation}
    \label{eqn:cfm_objective}
    \mathcal{L}_{CFM} = -\frac{1}{2} \mathbb{E}_{t \sim \mathcal{U}(t), \epsilon \sim \mathcal{N}(0, I)}[w_t \lambda_t'||\epsilon_{\theta}(x_t, t) - \epsilon||^2].
\end{equation}
\(\lambda_t'\) denotes the redefined signal-to-noise ratio and \(w_t\) represents a weighting factor that varies with time.

\subsubsection{Multimodal Attention.}
The DiT model~\cite{peebles2023scalable} is used in designs such as FLUX.1~\cite{flux2024}, Stable Diffusion 3~\cite{esser2024scaling}, and PixArt~\cite{chen2023pixart} iteratively refine noise image tokens using a transformer denoising network.
The DiT architecture operates on two token categories, namely noise image tokens ${X}\in \mathbb{R}^{N \times d}$ and text conditioning tokens ${C}_{\text{T}}\in \mathbb{R}^{M \times d}$. Here $d$ indicates embedding dimensionality, with $N$ and $M$ representing the respective counts of visual and textual tokens.
Throughout the network, these tokens maintain consistent shapes as they pass through multiple transformer blocks.
FLUX.1's DiT blocks include layer normalization, multimodal attention (MMA)~\cite{pan2020multi}, and rotary position embedding (RoPE)~\cite{su2024roformer} to encode spatial information.

Next, the multimodal attention module takes the tokens with positional encodings as its input and generates distinct query $Q$, key $K$, and value $V$ representations.
It facilitates the calculation of attention among all tokens:
\begin{equation}
  \text{MMA}([{X};{C}_{\text{T}}]) = \text{softmax}\left(\frac{QK^{\top}}{\sqrt{d}}
  \right)V, \label{eq:mma}
\end{equation}
where $[{X};{C}_{\text{T}}]$ represents the concatenation of image and text tokens. This concept supports bidirectional attention.
Building on the DiT architecture with FLUX.1 as the implementation basis, we aim to develop DealMaTe, a framework that balances superior performance with minimalist design, guiding material transfer exclusively via visual conditions.

\textcolor{black}{
\subsection{Multi-Dim 3D Shader LoRA}
\label{sec:3.2}
To efficiently integrate 3D structural conditional signals while preserving the generalization ability of the pre-trained model and the material representation capability of the diffusion backbone, we extend and optimize the FLUX architecture~\cite{flux2024}. Instead of adopting the traditional approach of adding independent control modules, which often directly embeds control modules into the main network layer and can interfere with the model original feature representation, as seen in the method proposed by Tan et al.~\cite{tan2025ominicontrol}, we introduce an additional conditional branch. This allows for the seamless and natural integration of 3D structural conditional information into the existing architecture.
Moreover, we deploy a dedicated Multi-Dim 3D Shader LoRA only on this new conditional branch. This strategy not only achieves the intended functionality but also avoids the generation of redundant parameters and the increase in additional computational costs. 
This separation is crucial, as it ensures precise structural control while preventing degradation of the model's core generative capabilities, and this key advantage ultimately enables the high-fidelity synthesis of both structure and material.
}

\input{Figures/attention}

\textcolor{black}{
In the Transformer architecture, the input features are first mapped to three types of features: queries (\( Q \)), keys (\( K \)), and values (\( V \)), which are then processed by the self-attention mechanism. When the input feature representations correspond to material, noise, and 3D shader conditions, respectively, their corresponding input representations are \( Z_m \), \( Z_n \), and \( Z_c \). The standard query, key, and value transformations can be defined as:
\begin{equation}
Q_i, K_i, V_i = W_Q Z_i, W_K Z_i, W_V Z_i, \quad i \in \{m, n, c\}.
\end{equation}
The projection matrices \(W_Q, W_K, W_V\) are shared between branches to share parameters, however this approach does not maximize the representation of conditional signals. To solve this constraint, we creatively insert the LoRA into the conditional path to adaptively boost the expressiveness of conditional features while guaranteeing that the parameters and structures of the other branches are unaffected:
\begin{equation}
\Delta Q_c = B_Q A_Q Z_c, \quad \Delta K_c = B_K A_K Z_c, \quad \Delta V_c = B_V A_V Z_c.
\end{equation}
In the low-rank adaptation matrices, \( A_Q, A_K, A_V \in \mathbb{R}^{r \times d} \) and \( B_Q, \\B_K, B_V \in \mathbb{R}^{d \times r} \), where \( r \ll d \). Under this setting, the query, key, and value in the conditional branch are updated to:
\begin{equation}
Q'_c = Q_c + \Delta Q_c, \quad K'_c = K_c + \Delta K_c, \quad V'_c = V_c + \Delta V_c.
\end{equation}
Meanwhile, the material and noise branches remain unchanged:
\begin{equation}
Q'_i = Q_i, \quad K'_i = K_i, \quad V'_i = V_i, \quad i \in \{m, n\}.
\end{equation}
}

\textcolor{black}{
By applying LoRA-based adaptation solely to the 3D shader conditional branch, we ensure that the conditional signal is efficiently integrated into the model without interfering with the material representation. This focused adjustment allows the model to flexibly incorporate conditional information while preserving the integrity of its original feature space, thereby enabling more controllable and high-fidelity material transfer, as demonstrated in Sec.~\ref{sec:4.5.1 3D-SCB}.
}

\textcolor{black}{
\subsection{Shader Casual Multimodal Attention}
\label{sec:3.3}
\subsubsection{Shader Casual Attention.}
Shader Casual Attention is a unidirectional attention mechanism whose core objective is to ensure strict adherence to temporal causality by restricting the flow of information within sequence models. Specifically, it stipulates that each position in the sequence can only attend to itself and the preceding positions. 
This rule is achieved by applying a mask composed of $0$ and $-\infty$ to the attention logits before the softmax operation, with the following mathematical expression:
\begin{equation}
\mathbf{Q} = [Q'_m; Q'_n; Q'_c],
\mathbf{K} = [K'_m; K'_n; K'_c],
\mathbf{V} = [V'_m; V'_n; V'_c],
\end{equation}
\begin{equation}
\text{MMA}([N;M;{C}_{\text{c}}]) =\textit{Softmax}(\mathbf{Q}\mathbf{K}^\top / \sqrt{d_k} + M) \mathbf{V}.
\end{equation}
In this mechanism, $M$ is responsible for ensuring the adherence to causality, while QKV is composed of features from the material, noise, and 3D condition branches. 
To improve efficiency and integrate multiple condition signals during inference, we have developed a new causal attention mechanism called Shader Casual Mutual Attention (SCMA). 
By employing different masking strategies, this mechanism can precisely control the flow of information, ensuring that while multiple condition signals are integrated, the independence between each condition signal is maintained. 
}

\subsubsection{Shader Causal Mutual Attention (SCMA)}
However, during multi-condition inference, the interaction between condition tokens and denoising tokens stays normal, but interference across conditions occurs due to the model's lack of training on cross-condition token interactions. The following mechanism, with its unique architecture, intelligently combines several condition signals while avoiding interference during the inference phase. We specifically describe the input sequence for multi-condition inference as follows:
\begin{equation}
Z = [Z_{\text{n}}; Z_{\text{m}}; Z_{\text{c}_{1}}; Z_{\text{c}_{2}}; Z_{\text{c}_{3}}].
\end{equation} 
In this mechanism, \( Z_{\text{n}} \) and \( Z_{\text{m}} \) are used to identify the tokens related to noise and material, respectively.
While \( Z_{\text{c}_1} \), \( Z_{\text{c}_2} \), and \( Z_{\text{c}_3} \) correspond to the depth, normal, and lighting tokens, respectively.

To precisely control the flow of attention, we introduce a special attention mask \( M \in \{0, -\infty\}^{n \times n} \), whose specific construction is as follows:
\begin{equation}
M_{ij} =
\begin{cases}
-\infty, & \text{if } i \notin n_{\text{t\&n}} \text{and} j\in n_{\text{t\&n}},  \\
0, & \text{otherwise}.
\end{cases}
\end{equation}  
Under this setup, the total sequence length is given by \( n = n_{\text{m\&n}} + \sum_{i=1}^{m} n_{\text{c}_i} \). This carefully designed masking strategy ensures that while the image tokens aggregate information from all conditions, the distinct conditions remain isolated from one another, maintaining their independence.
We present experimental validation in Sec.~\ref{sec:4.5.2 SCMA}.

\subsection{Training and Inference}
\label{sec:3.4}
\subsubsection{Loss Function}
The flow-matching loss is used as our loss function, defined as
\begin{equation}
L_{RF}=E_{t,\epsilon \sim N(0, I)}| | v_\theta(z, t, c_i)-\left(\epsilon-x_0\right)| |_2^2.
\end{equation}
In this formula, \( \epsilon \) represents the predicted noise, \( v_\theta \) denotes the velocity field, \( z \) is the image feature at time \( t \), \( c_i \) indicates the input condition, and \( x_0 \) is the original image feature.

\subsubsection{Efficient Inference via KV Cache.}
\label{kvcache}
To achieve efficient inference, we employ a novel key-value (KV) caching strategy specifically designed for multi-conditional diffusion models. Unlike previous approaches in diffusion inference optimization~\cite{wimbauer2024cache,ma2024learning} that primarily focus on reducing sampling steps or model distillation, our method focuses on decoupling conditional feature computation.
The core of this approach is to design the conditioning branch as a computation module independent of the denoising timesteps, leveraging the properties of the causal attention mechanism to make the conditioning branch agnostic to the timesteps.

During inference, we calculate and store the key-value pairs of all conditional features in the cache dictionary \(\mathcal{D}\) just once at the initial timestep, as the calculation of the conditioning branch is independent of individual timesteps. These cached KV pairs are subsequently utilized in all following denoising processes, preventing duplicate computation of similar conditional features.
Specifically, for \( 3 \) conditional features \(\{\text{cond}_i\}_{i=1}^3\), we compute at the initial timestep:
\begin{equation}
K_{C_i}, V_{C_i} = f_{\theta}(\text{cond}_i), \quad \forall i \in \{1, 2, 3\},
\end{equation}
where \(\text{cond}_1\), \(\text{cond}_2\), and \(\text{cond}_3\) correspond to depth, normals, and lighting conditions, respectively.
We store these in the cache \(\mathcal{D}[i] \gets (K_{C_i}, V_{C_i})\). In each subsequent denoising step \( t \), we directly retrieve the cached key-value pairs \(\{(K_{C_i}, V_{C_i})\}_{i=1}^3\), and compute the query, key, and value vectors corresponding to the current noise and text features:
\begin{equation}
Q_{\text{denoising}}, K_{\text{denoising}}, V_{\text{denoising}} = f_{\theta}(\text{x}_t, t).
\end{equation}
Subsequently, we concatenate the cached conditional key-value pairs with the current computed key-value pairs:
\begin{equation}
\begin{aligned}
Q &= Q_{\text{denoising}}, \\
K &= \text{Concat}\left(K_{\text{denoising}}, K_{C_1}, K_{C_2}, K_{C_3}\right), \\
V &= \text{Concat}\left(V_{\text{denoising}}, V_{C_1}, V_{C_2}, V_{C_3}\right).
\end{aligned}
\end{equation}
The attention output is then computed using the standard attention mechanism. This method significantly reduces inference latency by eliminating \( N \)-fold redundant computations (corresponding to \( N \) denoising steps), while maintaining generation quality and model flexibility. See Sec.~\ref{sec:4.5.6 KV Cache} for relevant experimental verification.

\newcommand{\img}{\mathbf{x}}
\newcommand{\depth}{\mathbf{d}}
\newcommand{\latent}{\mathbf{z}}
\newcommand{\latentdepth}{\latent^{(\depth)}}
\newcommand{\latentimage}{\latent^{(\img)}}
\newcommand{\noise}{\bm{\epsilon}}
\newcommand{\denoiser}{\bm{\epsilon}_{\theta}}
\newcommand{\denoiserlong}{\denoiser(\latentdepth_t, \latentimage, t)}
\newcommand{\catinput}{\mathbf{z}}
\newcommand{\encoder}{\mathcal{E}}
\newcommand{\decoder}{\mathcal{D}}

\newcommand{\pred}{\mathbf{\hat{d}}}
\newcommand{\translated}{\mathbf{\hat{d^{\prime}}}}
\newcommand{\merged}{\mathbf{m}}

\input{Figures/comparison}

\input{Tables/quantity}
\input{Figures/ablation_3D-SCB}

\subsection{Dataset Collection}
\label{sec:3.5}
For the spatial control LoRA training of depth, lighting, and normals, we primarily utilized 2,400 natural images from Unsplash~\cite{unsplash} that encompass arbitrary subject information. The corresponding depth, lighting, and normals for these natural images were obtained through Marigold~\cite{ke2023repurposing,ke2025marigold}. 
All of the images are preprocessed, including cropping and alignment, to guarantee that our training inputs are consistent and accurate.
This resulted in four subsets of data, each containing corresponding real, depth, lighting, and normal images.

\noindent\textbf{Depth.}
Previous material transfer methods~\cite{cheng2025marble,garifullin2025materialfusion,cheng2024zest} predominantly use depth as the structural condition, and we adopt and further strengthen this paradigm.
Given an image \(\img\), we use the Stable Diffusion VAE to encode it into a latent code \(\latentimage\). We then concatenate it with the depth latent code \(\latentdepth_t\) and input it to the fine-tuned U-Net for iterative denoising. After \(T\) steps of the schedule, the depth latent code \(\latentdepth_0\) is decoded into an image, and the average of its 3 channels is taken to obtain the initial depth estimate \(\hat\depth\).

To improve the quality of the depth-guiding results, we propose a novel test-time ensemble scheme that leverages the stochasticity of the inference process. For each input, we run \(N\) inferences to obtain a set of predictions \(\{\pred_1, \ldots, \pred_N\}\). We align these predictions by optimizing for the scales \(\hat{s}_i\) and offsets \(\hat{t}_i\), with the objective function given by:
\begin{equation}
\min_{\substack{s_1,\ldots,s_N \\ t_1, \ldots, t_N}} \Bigg( \sqrt{\frac{1}{b} \sum_{i=1}^{N-1} \sum_{j=i+1}^{N} \| \translated_i - \translated_j \|_2^2} + \lambda \mathcal{R} \Bigg),
\end{equation}
where \(\translated = \pred \times \hat{s} + \hat{t}\), \(\mathcal{R} = |\! \min (\merged)| + |1 - \max (\merged)|\), and \(b = \binom{N}{2}\). During optimization, the merged depth \(\merged\) is computed via pixel-wise median, and it serves as the final ensemble prediction. This scheme does not require real data and balances efficiency and quality by adjusting \(N\).

\noindent\textbf{Normals.}
Surface normals are closely related to depth estimation, as both aim to recover 3D geometric information. Depth estimation outputs scalar values, while normals are represented as three-dimensional unit vectors. 
Real normal annotations are difficult to collect and are often generated from depth, which might result in noise in flat areas and excessive smoothing at edges.
Marigold-Normals bridges the gap between simulation and reality through the Stable Diffusion prior.
We first run inference \(N\) times with different noise initializations. The resulting normal predictions \(\{\mathbf{\hat{n}}_1, \ldots, \mathbf{\hat{n}}_N\}\) are averaged into \(\mathbf{\bar{n}}\) and normalized. Finally, for each pixel \((u,v)\), we select the prediction with the highest cosine similarity to the mean normal \(\mathbf{\bar{n}}^{(u,v)}\):
\begin{equation}
\operatorname*{arg\,max}_{i} \, \mathbf{\bar{n}}^{(u,v)} \cdot \mathbf{\hat{n}}^{(u,v)}_i.
\end{equation}

\noindent\textbf{Lighting.}
We obtain the lighting image using Marigold-IID-Lighting by first feeding the input image \( I \) into the model, which outputs three decomposition components: albedo \( A \), diffuse shading \( l \), and non-diffuse residual \( R \). The diffuse shading component \( l \) corresponds directly to the lighting image, representing illumination color. The decomposition follows the formula:
\begin{equation}
I = A \cdot l + R.
\end{equation}
By extracting the predicted \( l \) component from the model, we obtain the lighting image. 
This formulation gives an organized approach to distinguishing between reflectance and shading while taking into consideration complicated lighting processes.

%% file: Figures/attention.tex
\begin{figure}
\centering
\includegraphics[width=\linewidth]{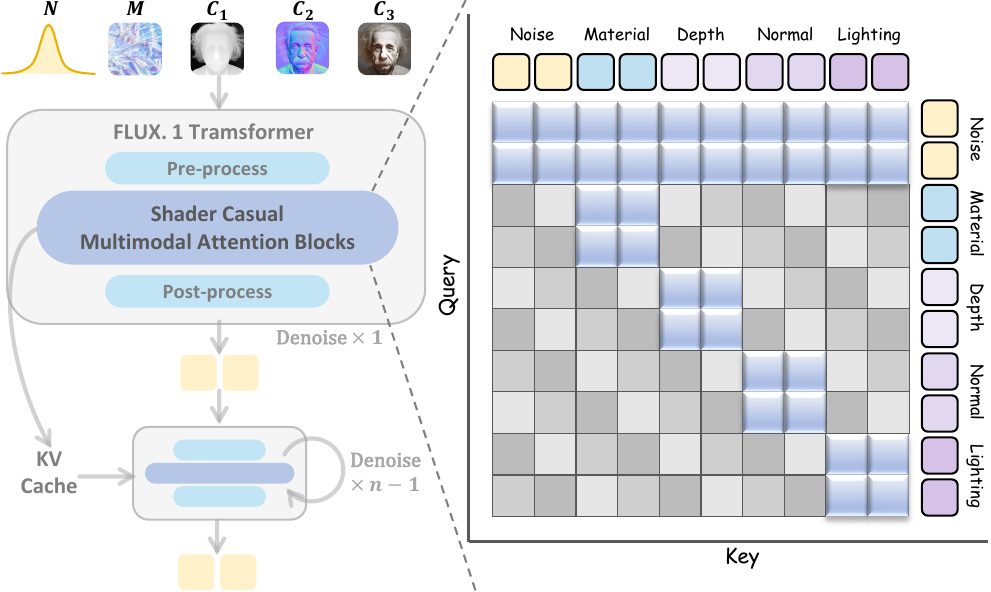}
\caption{To optimize the inference performance of the material transfer task, the framework is novelly designed with a shader causal multimodal attention mechanism. This mechanism aligns with the task's requirements for feature correlation processing. Besides, it can directly enable key-value caching, ultimately reducing inference latency and improving efficiency effectively.}
\label{fig:attention}
\end{figure}

%% file: Figures/comparison.tex
\begin{figure*}
\centering
\includegraphics[width= \linewidth]{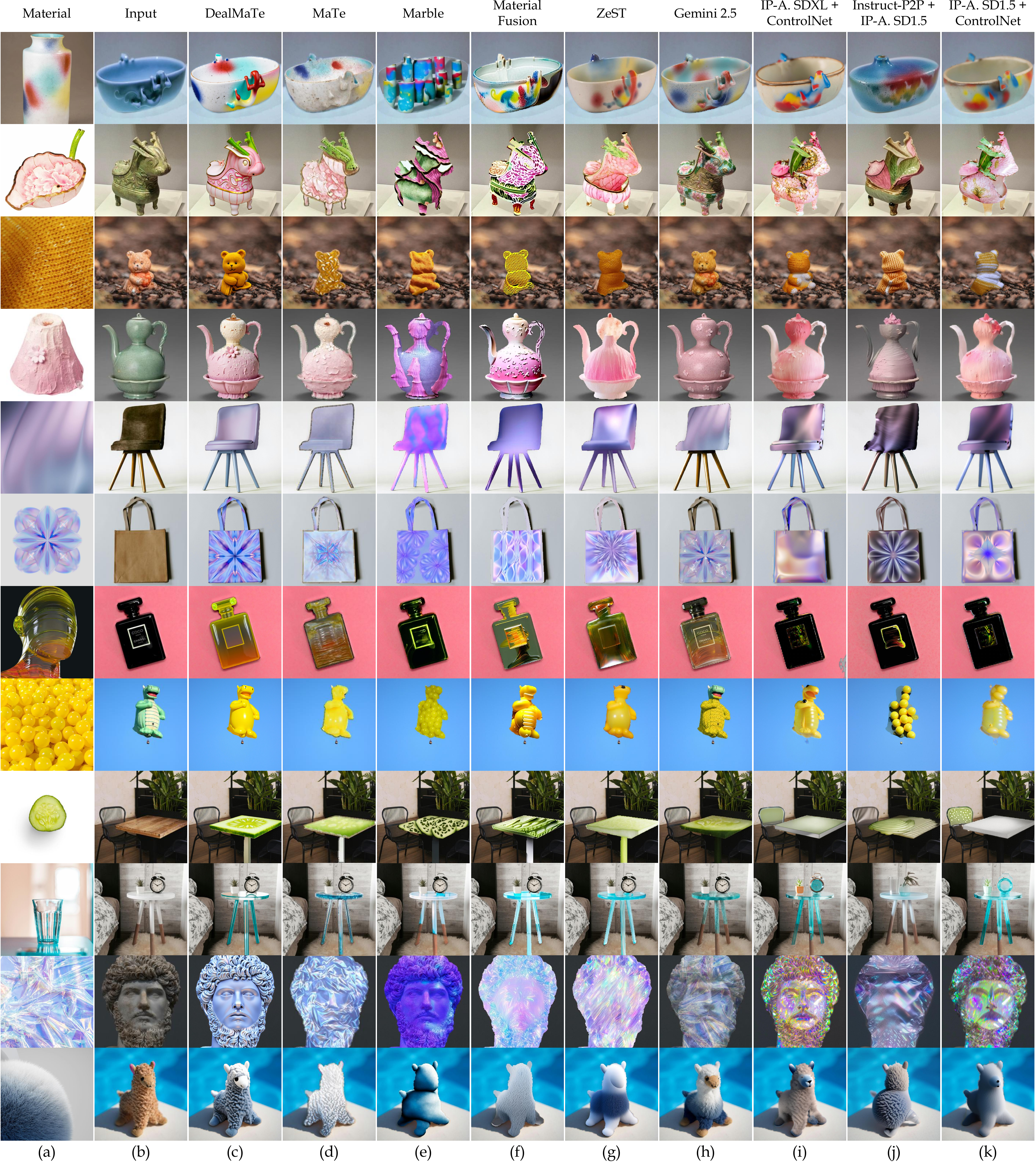}
\vspace{-6mm}
\caption{
Qualitative comparisons with other methods. Column (d) shows the result of our conference version work MaTe. Columns (e)-(g) show a qualitative comparison of DealMaTe with several state-of-the-art material transfer methods. Column (h) is one of the most advanced image generation agents that can understand complex editing commands. Columns (i)-(k) show the pipeline we build using the IP-Adapter and reference networks.}
\label{fig:comparison}
\end{figure*}

%% file: Tables/quantity.tex
\begin{table*}
\centering
\caption{Quantitative comparisons with other methods are presented. The results that represent the current SOTA baselines for material transfer are highlighted in gray, with the best results shown in bold and the second-best results underlined.}
\setlength{\tabcolsep}{3.5pt} 
\begin{tabular}{c||c|cccccccc}
\toprule
& \cellcolor{LightBlue}DealMaTe & \cellcolor{gray!15}\makecell{MaTe\\ \textit{[ICCV 2025]}} & \cellcolor{gray!15}\makecell{Marble\\ \textit{[CVPR 2025]}} & \cellcolor{gray!15}\makecell{Material\\Fusion} & \cellcolor{gray!15}\makecell{ZeST\\\textit{[ECCV 2024]}} & Gemini 2.5 & \makecell{IP-A. SDXL \\+ ControlNet} & \makecell{Inst-P2P\\+ IP-Adapter} & \makecell{IP-A. SD1.5\\+ ControlNet} \\ \hline\hline

\textbf{SSIM $\uparrow$} & \cellcolor{LightBlue}\textbf{0.8906} & \cellcolor{gray!15}0.8176 & \cellcolor{gray!15}0.7038 & \cellcolor{gray!15}0.8263 & \cellcolor{gray!15}0.7231 & \underline{0.8825} & 0.8152 & 0.7093 & 0.7489 \\ \hline

\textbf{LPIPS $\downarrow$} & \cellcolor{LightBlue}\textbf{0.1285} & \cellcolor{gray!15}0.1548 & \cellcolor{gray!15}0.2970 & \cellcolor{gray!15}0.1565 & \cellcolor{gray!15}0.2430 & \underline{0.1317} & 0.1876 & 0.2743 & 0.2537 \\ \hline

\textbf{CLIP $\uparrow$} & \cellcolor{LightBlue}\textbf{0.8927} & \cellcolor{gray!15}\underline{0.8834} & \cellcolor{gray!15}0.7816 & \cellcolor{gray!15}0.8521 & \cellcolor{gray!15}0.8627 & 0.8692 & 0.8358 & 0.7904 & 0.8271 \\ \hline

\textbf{PSNR $\uparrow$} & \cellcolor{LightBlue}\textbf{16.755} & \cellcolor{gray!15}14.987 & \cellcolor{gray!15}15.337 & \cellcolor{gray!15}\underline{16.308} & \cellcolor{gray!15}15.540 & 15.0323 & 9.4670 & 11.482 & 8.9032 \\ \hline

\textbf{DreamSim $\downarrow$} & \cellcolor{LightBlue}\textbf{0.3527} & \cellcolor{gray!15}0.3841 & \cellcolor{gray!15}0.4352 & \cellcolor{gray!15}0.4038 & \cellcolor{gray!15}0.6020 & \underline{0.3832} & 0.6089 & 0.6897 & 0.6676 \\
\bottomrule
\end{tabular}
\label{tab:comparison}
\end{table*}

%% file: Figures/ablation_3D-SCB.tex
\begin{figure}
\centering
\includegraphics[width=\linewidth]{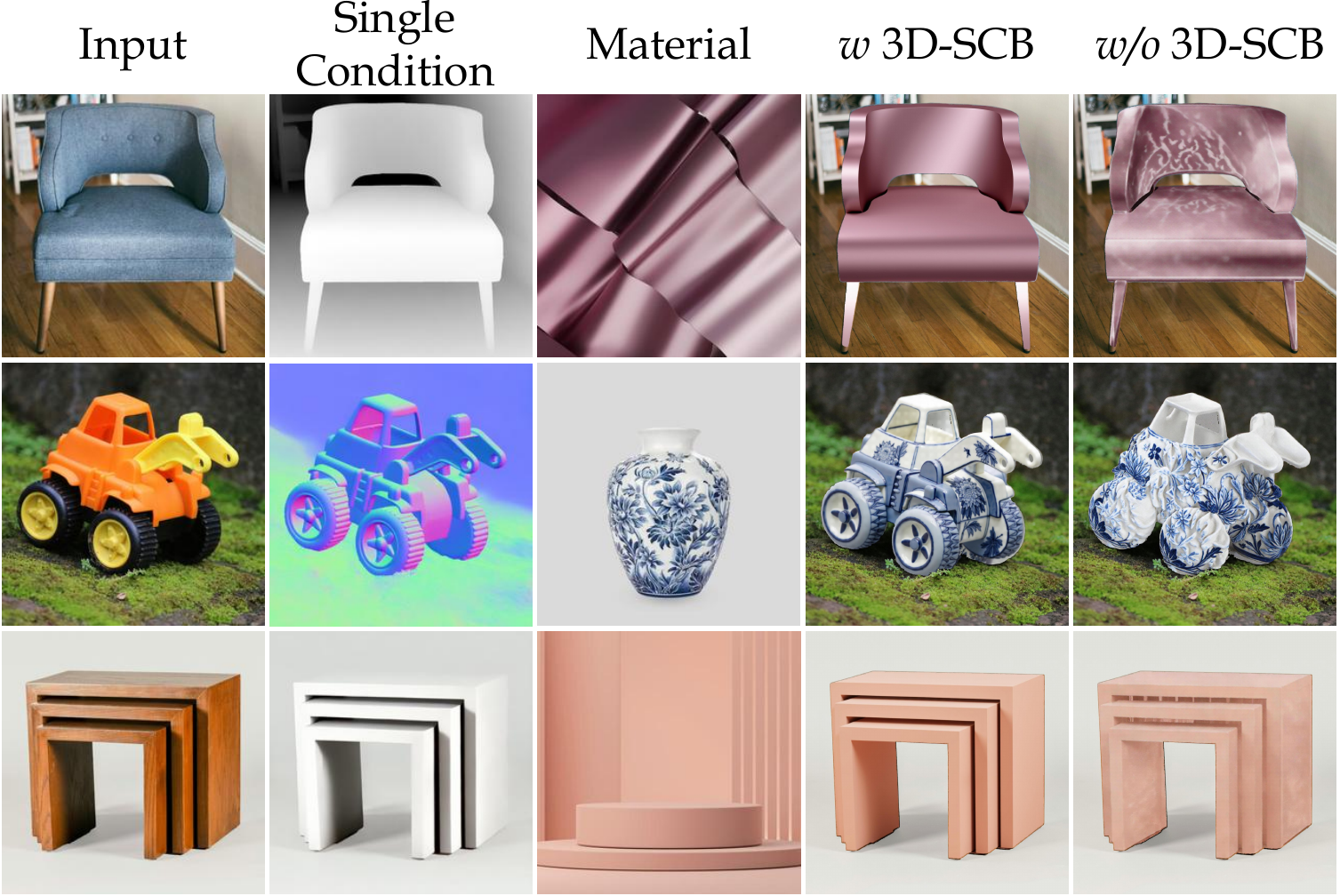}
\caption{When 3D-SCB is removed, the surface textures become contaminated with artifacts and irregular bumps, and naive condition interference becomes evident.}
\label{fig:ablation_3D-SCB}
\end{figure}

%% file: Sections/4_experiments.tex
\section{Experiments}
\subsection{Implementation Details}
%
We base the method on FLUX.1~\cite{flux2024}, a latent rectified flow transformer for image generation. 
In our experiments, we set the inference steps to $25$ and the image generation size to $1024 \times 1024$ pixels. All the inference experiments are conducted on a single NVIDIA $A100$ GPU. Each 3D Shader LoRA takes roughly three days to train for $500k$ steps on $4$ NVIDIA $A100$ GPUs. 
We employ a batch size of one and a learning rate of $1\cdot 10^{-4}$.

For comparison experiments settings, we compare DealMaTe with Marble~\cite{cheng2025marble}, Material Fusion~\cite{garifullin2025materialfusion}, ZeST~\cite{cheng2024zest}, Gemini 2.5~\cite{comanici2025gemini},  IP-Adapter SDXL~\cite{ye2023ip-adapter}, Instruct-P2P~\cite{brooks2023instructpix2pix}, and IP-Adapter SD1.5~\cite{ye2023ip-adapter} in the task of material transfer.
To enhance the fairness of the experiments, we added the corresponding version of the ControlNet depth control models~\cite{control_v11f1p_sd15_depth,controlnet-depth-sdxl-1.0} to the pipelines for IP-Adapter SD1.5 and IP-Adapter SDXL. Additionally, we incorporated mask segmentation operations for Gemini 2.5, IP-Adapter SDXL / SD 1.5 + ControlNet, and Instruct-P2P + IP-Adapter SD 1.5.

\input{Figures/ablation_SCMA}

\input{Figures/ablation_wo_one_condition}
\input{Tables/userstudy}

\subsection{Qualitative Comparison}
As shown in Fig.~\ref{fig:comparison}, our method excels in preserving the structure of the input images, faithfully rendering the structural characteristics of the original images. Material transfer based solely on image conditions has been unattainable with previous methods~\cite{cheng2025marble,garifullin2025materialfusion,cheng2024zest}. In contrast, the DealMaTe method proposed in this paper (see Fig.~\ref{fig:comparison}~(c)) achieves balanced alignment between structural preservation and material representation by optimizing the extraction mechanism of appearance information from material images and accurately capturing the mapping relationship between the subject structure and the target material. 
These results demonstrate that our method makes a meaningful step forward in tackling current challenges, offering an effective new paradigm for material transfer studies.

When compared to our conference version MaTe~\cite{Huang_2025_ICCV} (see Fig.~\ref{fig:comparison}~(d)), DealMaTe demonstrates significant qualitative improvements in maintaining rigorous geometric consistency. While the baseline MaTe achieves general texture alterations, it frequently struggles with structural fidelity and fine-grained object details, issues that our enhanced architecture successfully resolves.
Moreover, methods specifically designed for material transfer tasks, such as Marble, Material Fusion, and ZeST, have significant shortcomings in preserving structural information. 
As shown in Fig.~\ref{fig:comparison}~(e)-(g), these methods often struggle to effectively retain such features, whether it is the morphological details of antiques, the contour features of toys, or the structural information of human limbs and faces.
This further indicates that existing advanced material transfer methods still face technical bottlenecks in preserving the 3D structural information of complex objects.

From the comparative experimental results, it is also evident that the mainstream Gemini 2.5 method, despite its representative performance, fails to completely strip away the original features of the input images, resulting in incomplete material transfer (see Fig.~\ref{fig:comparison}~(h)). Taking the ``toy bear'' as an example, the result processed by this method shows almost no noticeable change in material properties.
Furthermore, to verify the effectiveness of the method, this paper also constructed three types of composite processing pipelines for supplementary experiments (see Fig.~\ref{fig:comparison}~(i)-(k)).
Among them, the combination of ControlNet and IP-Adapter (based on SDXL or SD1.5) has improved the overall structural expression, but the preservation effect of key facial details such as eyes, nose, and mouth is still not satisfactory. On the other hand, the combination of Instruct-P2P and IP-Adapter shows a certain uniqueness in material reproduction but has significant deviations in the accuracy of structural information.

\input{Figures/ablation_depth}
\input{Figures/ablation_normal}

\subsection{Quantitative Comparison}
We quantitatively validate the approach on the MTB benchmark~\cite{Huang_2025_ICCV} with regard to state-of-the-art material transfer methods.  
The Structural Similarity (SSIM)~\cite{wang2004image} index and Peak Signal-to-Noise Ratio (PSNR)~\cite{psnr} are used to assess structural and geometric consistency between input images and outputs. 
The Learned Perceptual Image Patch Similarity (LPIPS)~\cite{zhang2018unreasonable} and the recent DreamSim~\cite{fu2023dreamsim} metric are used to evaluate the consistency of content detail and mid-level perceptual features between input and output images. We apply the Contrastive Language-Image Pre-training (CLIP)~\cite{radford2021learning} to evaluate the semantic similarity between the materials and the outcomes. 
Table~\ref{tab:comparison} provides a complete presentation of the experimental results. 
Our methodology consistently outperforms existing state-of-the-art approaches across evaluated metrics. Notably, compared to the conference version~\cite{Huang_2025_ICCV}, DealMaTe achieves significant performance gains due to its improved capability in disentangling complex material properties and enforcing geometric constraints. This shows that our model is highly effective and well-suited for tackling the specific task of material transfer.

\subsection{User Study}
Subjective evaluation consists of three main dimensions: material consistency, structure preservation, and overall quality.
Participants were asked to select their most preferred method from each dimension. We assembled a group of $19$ experts in the fields of vision and graphics to conduct the assessment. Through $30$ comparative experiments (each involving $8$ different methods), we collected a total of $1,710$ voting results from these three perspectives. The percentage data displayed in Table~\ref{tab:userstudy} reveals the preferences of the participants, demonstrating that our proposed DealMaTe method is more favored across all three dimensions.
By comparing the quantitative experimental results with the user study outcomes, we recognize the differences between objective and subjective evaluations. While objective evaluations typically consider each dimension independently, participants may synthesize information across dimensions when making their choices, even though they are presented with separate options.
The preferences of the participants indicate that our generated results achieve a better balance in material texture, structure preservation, and overall visual appeal. For more information, please see the supplementary materials.

\subsection{Ablation Study}
\subsubsection{3D shader conditional branch (3D-SCB)}
\label{sec:4.5.1 3D-SCB}
In our ablation study we analyze the impact of removing the 3D-SCB.  
Specifically, as illustrated by the `Single Condition' column in Fig.~\ref{fig:ablation_3D-SCB}, this ablation setting retains only a single condition signal and removes the independent LoRA structure of the conditional branch. Instead, the condition is directly injected into the main network using a naive LoRA structure. 
Without the 3D-SCB, the material details deviate from the source because such a generic condition LoRA injection corrupts the material latent space, as shown in the fifth column of Fig.~\ref{fig:ablation_3D-SCB}.
For metallic and plastic samples spurious patterns emerge on the surface, while for the blue-and-white porcelain unexpected bumps appear.  
The 3D-SCB preserves micro-textures while strictly adhering to depth, normal, and lighting constraints, delivering a faithful appearance.
This benefit is quantitatively confirmed in Table~\ref{tab:ablation_quantity} where the CLIP score of 0.8828 is rendered in the deepest purple in its column, underscoring the positive role of 3D-SCB for material fidelity.

\input{Figures/ablation_lighting}

\subsubsection{Shader Casual Mutual Attention (SCMA)}
\label{sec:4.5.2 SCMA}
We conduct an ablation study by removing the SCMA module and replacing it with standard multimodal attention (MMA, as formulated in Eq.~\ref{eq:mma}) to evaluate its contribution. As illustrated in Fig.~\ref{fig:ablation_SCMA}, the replacement of SCMA with standard MMA disables the independent guidance from different structural conditions and results in negligible structural refinement. 
Quantitative results in Table~\ref{tab:ablation_quantity} confirm that eliminating SCMA leads to clear degradation in both SSIM and LPIPS. These findings demonstrate that SCMA is essential for maintaining the independence and mutual non-interference of each structural condition, which jointly enables superior structural control.

\input{Tables/ablation_quantity}

\subsubsection{Single-Factor Control Ablation.}
To strictly verify the necessity of each spatial control signal in our conditional branch, we conducted a leave-one-out ablation study, as shown in Fig.~\ref{fig:ablation_wo_one_condition}. In this experiment, we remove exactly one input condition to observe the specific degradation in the generated results.
First, without depth guidance, the depth of field and separation between the foreground and background are imprecise, leading to a lack of 3D volume. Second, without lighting guidance, the generated objects appear extremely flat, and critical semantic details, such as the clear structure of the dog’s eyes and nose, are severely degraded or washed out. Finally, without normal guidance, the model struggles to map complex materials accurately onto uneven surfaces, causing severe texture discontinuities and a loss of fine-grained structural fidelity. Overall, these ablation results show that depth, lighting, and normal control different and complementary spatial properties. Removing any single factor harms the structural consistency of material transfer.

\subsubsection{Multi-Dim 3D Shader LoRA Effect Strength.}
The effects of each LoRA's intensity on transfer quality are shown in Figs.~\ref{fig:ablation_depth}, \ref{fig:ablation_normal}, and \ref{fig:ablation_lighting}. 
In practice, setting these parameters involves a general trade-off between rigidly adhering to the target object's spatial and lighting constraints and preserving the source material's intrinsic appearance.
For the depth LoRA ($\lambda$), lower strength ($\lambda < 1.0$) helps retain the material's inherent characteristics, such as surface undulations and micro-texture details, but may result in insufficient adherence to the global geometric structure. When $\lambda \ge 1.0$, the depth constraint is fully activated, ensuring the material accurately conforms to the target object's contours. However, pushing beyond this threshold yields no further structural improvement and risks overly constraining the material's natural displacement, leading to rigid appearance.
Similarly, the normal LoRA ($\gamma$) balances surface curvature alignment against the material's inherent geometric features. Lower strength ($\gamma < 1.2$) better preserves the material's original physical texture and gloss expression but fails to wrap the material accurately around the object's curved surfaces. At $\gamma = 1.2$, harmonious integration is achieved by fitting the target's surface normals without completely flattening the material's intrinsic appearance. Yet higher strength may weaken the material-specific gloss and translucency characteristics.
For the lighting LoRA ($\mu$), lower strength helps preserve the material's base color and inherent reflective properties (such as glossiness and color purity) but fails to correct the lighting direction, resulting in inconsistent shading. At $\mu = 0.8$, a balance is struck between lighting consistency and material fidelity. Excessive strength ($\mu > 0.8$), while improving lighting consistency, leads to overexposure, negatively impacting accurate color presentation and potentially destroying fine optical effects such as translucency.
Therefore, to achieve an optimal balance among structure preservation, material fidelity, and lighting consistency, $\lambda$, $\gamma$, and $\mu$ are empirically set to 1.0, 1.2, and 0.8, respectively.

\input{Tables/ablation_kvcache}

\input{Figures/application}


\subsubsection{KV Cache.}
\label{sec:4.5.6 KV Cache}
As shown in Table~\ref{tab:ablation_kvcache}, the proposed KV cache mechanism significantly improves inference efficiency while maintaining output quality. Experimental results demonstrate that with 25 sampling steps, enabling KV cache reduces inference time from 73 seconds to 28 seconds, achieving a 2.61× speedup ratio. This confirms that our method effectively eliminates redundant computation of conditional features during the denoising process, thereby accelerating the generation procedure.
Notably, the performance improvement does not come at the cost of output quality degradation. The essentially unchanged CLIP score indicates that the visual quality and semantic alignment of the generated 1024$\times$1024 resolution images are preserved.

%% file: Figures/ablation_SCMA.tex
\begin{figure}
\centering
\includegraphics[width=\linewidth]{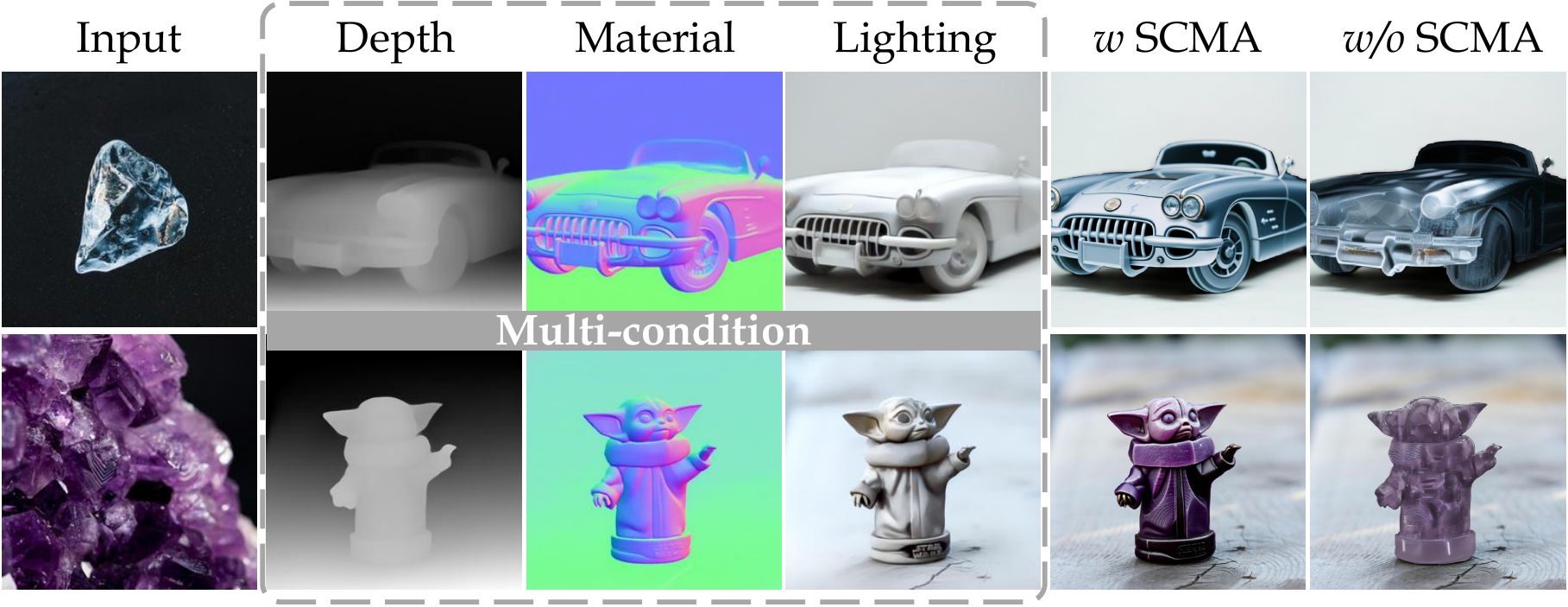}
\caption{Visual comparison of the SCMA ablation.
With SCMA, the generated structures remain clean; without it, structures are partly lost, and the different structural conditions mutually corrupt, yielding blurred details.}
\label{fig:ablation_SCMA}
\end{figure}

%% file: Figures/ablation_wo_one_condition.tex
\begin{figure*}
\centering
\includegraphics[width=\linewidth]{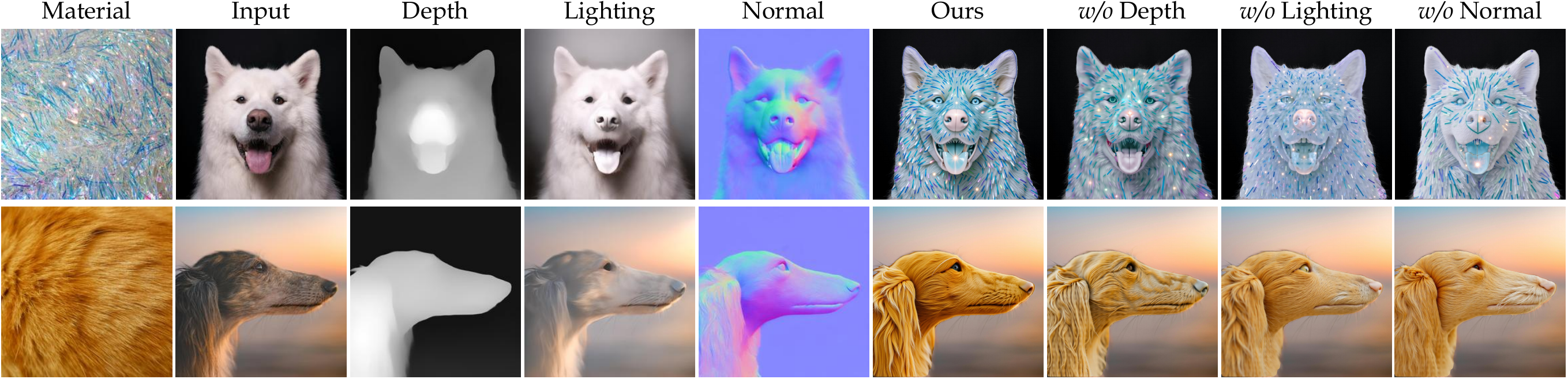}
\caption{We evaluate the necessity of each control signal by removing exactly one condition (depth, lighting, or normal) from our 3D Shader Conditional Branch.}
\label{fig:ablation_wo_one_condition}
\end{figure*}

%% file: Tables/userstudy.tex
\begin{table*}[!ht]
\centering
\caption{The user study results show the percentage of participants who preferred the outcomes of the corresponding methods in different aspects. The results that represent the current SOTA baselines for material transfer are highlighted in gray, with the best results shown in bold and the second-best results underlined.}
\begin{tabular}{c||c|cccccccc}
\toprule
& \cellcolor{LightBlue}DealMaTe  & \cellcolor{gray!15}\makecell{Marble\\ \textit{[CVPR 2025]}} & \cellcolor{gray!15}\makecell{Material\\Fusion} & \cellcolor{gray!15}\makecell{ZeST\\\textit{[ECCV 2024]}} & Gemini 2.5 & \makecell{IP-A. SDXL \\+ ControlNet}
 & \makecell{Instruct-P2P\\+ IP-Adapter} & \makecell{IP-A. SD1.5\\+ ControlNet} \\ \hline\hline
\textbf{Material $\uparrow$} & \cellcolor{LightBlue}\textbf{68.77\%} & \cellcolor{gray!10}0.88\% & \cellcolor{gray!10}2.63\% & \cellcolor{gray!10}6.49\% & \underline{9.82\%} & 8.42\% & 1.40\% & 1.58\% \\ \hline
\textbf{Structure $\uparrow$} & \cellcolor{LightBlue}\textbf{79.30\%} & \cellcolor{gray!10}0.70\% & \cellcolor{gray!10}2.28\% & \cellcolor{gray!10}1.75\% & \underline{7.54\%} & 4.39\% & 1.23\% & 2.81\% \\ \hline
\textbf{Overall $\uparrow$} & \cellcolor{LightBlue}\textbf{85.79\%} & \cellcolor{gray!10}0.53\% & \cellcolor{gray!10}1.75\% & \cellcolor{gray!10}1.93\% & \underline{4.91\%} & 2.98\% & 0.88\% & 1.23\% \\
\bottomrule
\end{tabular}
\label{tab:userstudy}
\end{table*}


%% file: Figures/ablation_depth.tex
\begin{figure*}[t]
\centering
\includegraphics[width= \linewidth]{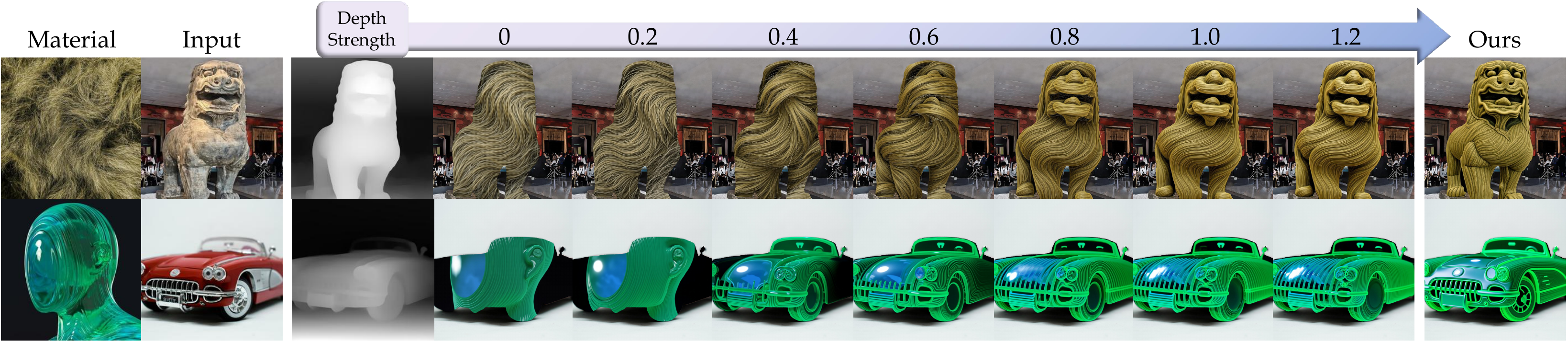}
\caption{Ablation experiment on the varying depth control parameter \(\lambda\).
}
\label{fig:ablation_depth}
\end{figure*}

%% file: Figures/ablation_normal.tex
\begin{figure*}
\centering
\includegraphics[width= \linewidth]{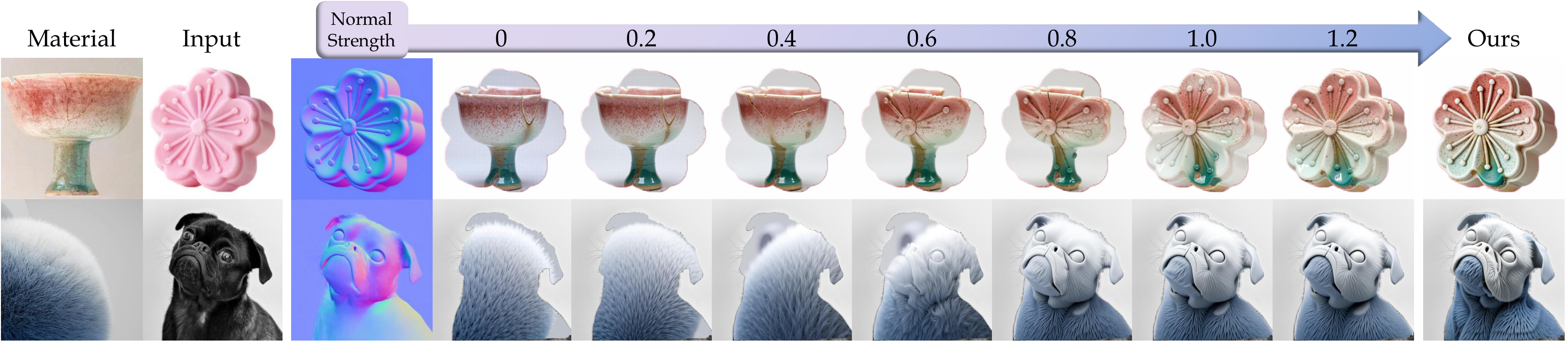}
\caption{Ablation analysis of the varying normal control parameter \(\gamma\).
}
\label{fig:ablation_normal}
\end{figure*}

%% file: Figures/ablation_lighting.tex
\begin{figure*}
\centering
\includegraphics[width= \linewidth]{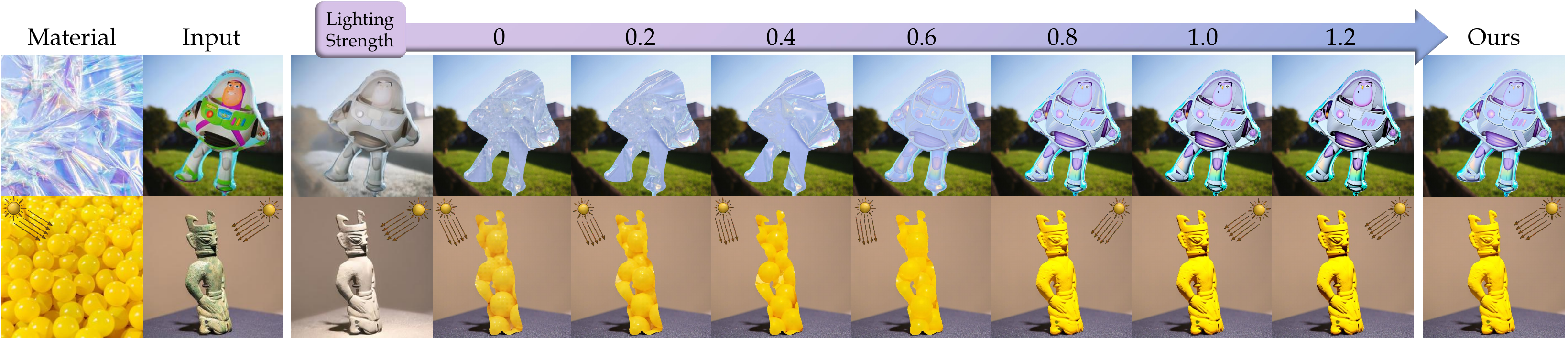}
\caption{Ablation experiment on the varying lighting control parameter \(\mu\).
}
\label{fig:ablation_lighting}
\end{figure*}

%% file: Tables/ablation_quantity.tex

\begin{table}
    \centering
    \caption{
    Ablation studies on 3D-SCB and SCMA on quantitative metrics.
    `Avg.' denotes the arithmetic mean across each ablation configuration.
    Cells outperforming the average are shaded \colorbox{teal!45}{teal}, while underperforming ones are shaded \colorbox{marron!45}{maroon}; darker hues indicate larger deviations.}
    \begin{tabular}{lccc}
    \toprule
     Method& \textbf{SSIM $\uparrow$} & \textbf{LPIPS $\downarrow$} & \textbf{CLIP $\uparrow$} \\
    \midrule
    \textit{Avg.} & \textit{0.8704} & \textit{0.1314} & \textit{0.8872} \\
    \midrule
    w/o 3D-SCB & \cellcolor{marron!25}0.8634 & \cellcolor{teal!15}0.1309 & \cellcolor{marron!80}0.8828 \\
    w/o SCMA & \cellcolor{marron!75}0.8572 & \cellcolor{marron!100}0.1348 & \cellcolor{marron!18}0.8862 \\
    Ours & \cellcolor{teal!55}0.8906 & \cellcolor{teal!45}0.1285 & \cellcolor{teal!45}0.8927 \\
    \bottomrule
    \end{tabular}

\label{tab:ablation_quantity}
\end{table}

%% file: Tables/ablation_kvcache.tex
\begin{table}[!t]
    \centering
    \caption{
    Ablation studies on KV cache. Calculation of the inference time is performed for images at 1024$\times$1024 resolution.}
    \resizebox{\linewidth}{!}
    {
    \begin{tabular}{lcccc}
    \toprule
     Method & \textbf{Sample steps} & \textbf{Inference time (s) $\downarrow$} & \textbf{Speedup $\uparrow$}& \textbf{CLIP Score $\uparrow$} \\
    \midrule
    w/o KV Cache & 25 & 73 & 1.00$\times$ & 0.8941\\
    w/ KV Cache & 25 & 28 & 2.61$\times$ & 0.8927\\
    \bottomrule
    \end{tabular}    }
\label{tab:ablation_kvcache}
\end{table}

%% file: Figures/application.tex
\begin{figure}
\centering
\includegraphics[width=\linewidth]{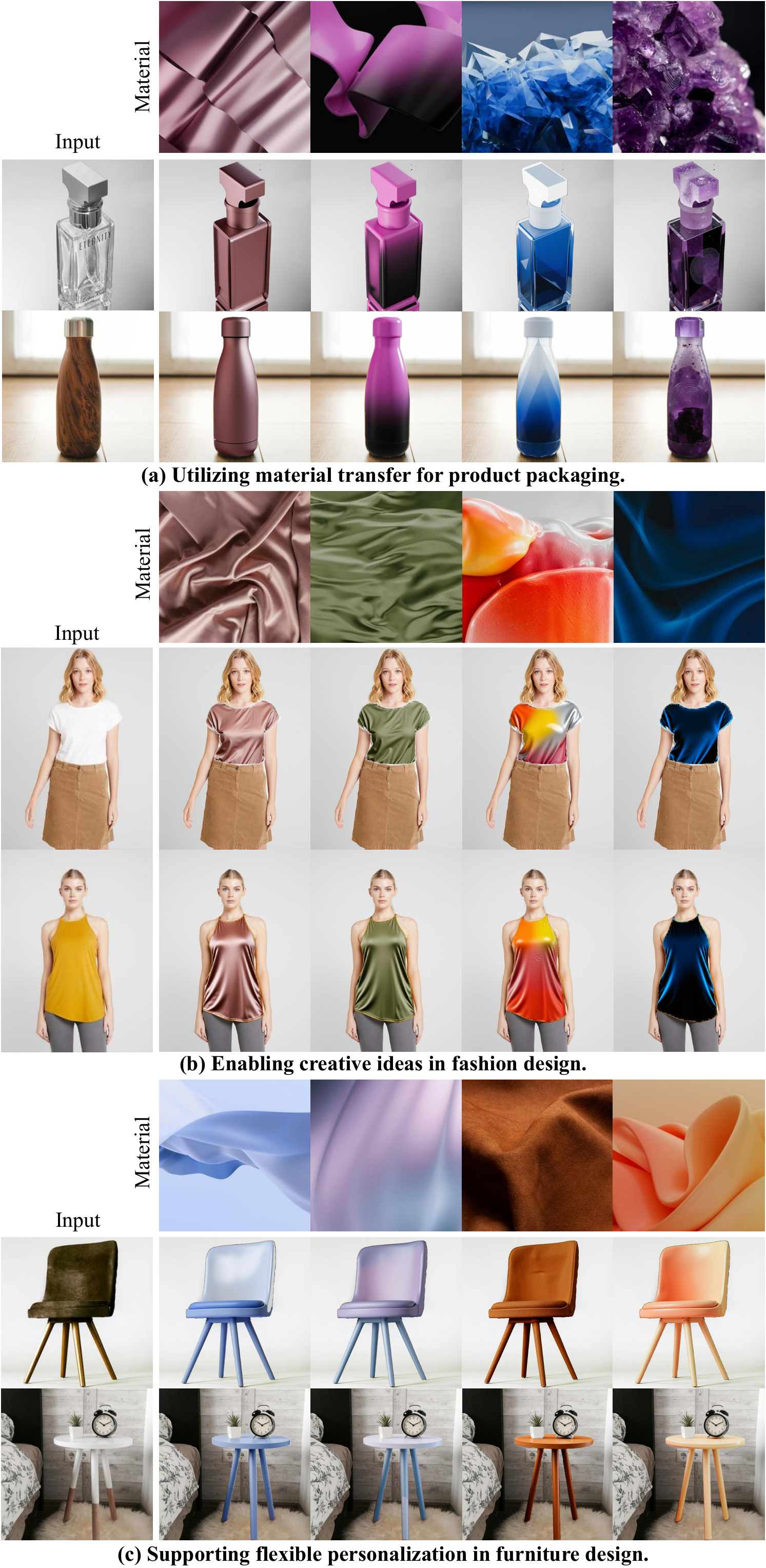}
\vspace{-0.5cm}
\caption{Various downstream applications of our material transfer method. From top to bottom: (a) product packaging, (b) garment fabric, and furniture design. Our approach demonstrates robust generalization and high-fidelity texture synthesis across diverse object categories and complex real-world scenarios.}
\vspace{-0.5cm}
\label{fig:application}
\end{figure}

%% file: Sections/5_Discussion.tex
\section{Applications and Discussions}
\subsection{Applications}
To verify the practicality and versatility of DealMaTe in material transfer, we applied it to fields such as product packaging, garment fabrics, and furniture design, demonstrating its adaptability and flexibility under various design requirements.

\noindent\textbf{Product packaging.}
In product packaging design, the visual effect of materials is crucial for attracting consumers. 
As shown in Fig.~\ref{fig:application}(a), our method can transfer the textures of metals or gemstones to packaging, achieving a high-end visual effect at a lower cost.
By adjusting model parameters, we can simulate the reflective and refractive properties of different materials, allowing for a preview of the final packaging effect during the design phase. This approach saves the cost of physical sample production and accelerates the design iteration process.

\noindent\textbf{Garment fabric.}
In the garment industry, the texture and color of fabrics are decisive for the final appearance of clothing. 
DealMaTe (see Fig.~\ref{fig:application}(b)) can transfer the smooth texture of silk to cotton fabric, giving it a high-end silk appearance visually.
This not only enhances the innovation in garment design but also allows for the prediction of the effects of different material combinations before production, reducing material waste and production costs.

\noindent\textbf{Furniture design.}
In furniture design, the choice of materials directly affects the durability and aesthetics of the furniture. Our model can transfer the natural texture of plastic to synthetic materials, giving them a natural plastic appearance visually. 
This not only enhances the aesthetic appeal of furniture design but also allows for the prediction of the effects of different material combinations before production, optimizing material selection and the manufacturing process, as shown in Fig.~\ref{fig:application}(c).

\input{Figures/failure_cases}
\input{Figures/failure_condition}

\subsection{Limitations and Discussions}
Although our proposed material transfer method generally achieves satisfactory results, it still has some limitations under specific conditions. 
First, we formally define a limitation called geometric-material mismatch. This conflict happens when a material requires a specific physical texture that contradicts the fixed shape of the input image. For instance, rendering matcha powder requires a granular and uneven physical structure. Our framework tightly restricts the generation process using the original depth and normal conditions. This restriction strongly suppresses the fine physical bumps needed for the powder effect. As Fig.~\ref{fig:failure_cases}~(a) illustrates, the model successfully transfers the green hue. Yet, it leaves the surface smooth and fails to capture the realistic physical nature of the powder.
Second, due to the nature of zero-shot learning, some detail features in the texture may be overemphasized, as shown in Fig.~\ref{fig:failure_cases}~(b).
%
Furthermore, the structural consistency of the results generated by DealMaTe is fundamentally bottlenecked by the accuracy of the underlying 3D geometric condition estimator~\cite{ke2025marigold}. As shown in Fig.~\ref{fig:failure_condition}, the estimator may fail to capture fine geometric changes, such as vertical (first row) or horizontal (second row) ridges, as well as complex ripple structures (third row). Therefore, the transferred results lack these complex structures and appear overly smooth or flattened in these areas. Addressing the robustness of pre-processing condition extraction for highly complex geometries remains an important direction for future research.

Additionally, if the material map contains multi-material information (as shown in Fig.~\ref{fig:discussion}(a)), multiple types of material elements are all transferred to the result. Finally, if the material includes particularly large special shapes (as illustrated in Fig.~\ref{fig:discussion}(b)), these shapes may also be transferred.
Furthermore, as shown in Fig.~\ref{fig:discussion}(c)), when the reference material is illuminated by colored light, the model tends to interpret this illumination color as an intrinsic property of the material, resulting in the colored lighting being transferred to the generated object.
To address these issues, future work will focus on optimizing the model's precision in identifying materials in images to better handle these special cases and improve the overall quality and reliability of material transfer.
Future work will focus on enhancing the model's accuracy in enabling region-aware or category-aware materials selection within one image to better handle these special cases and improve the overall quality and reliability of material transfer.

\input{Figures/discussion}

%% file: Figures/failure_cases.tex
\begin{figure}[t]
\centering
\includegraphics[width= \linewidth]{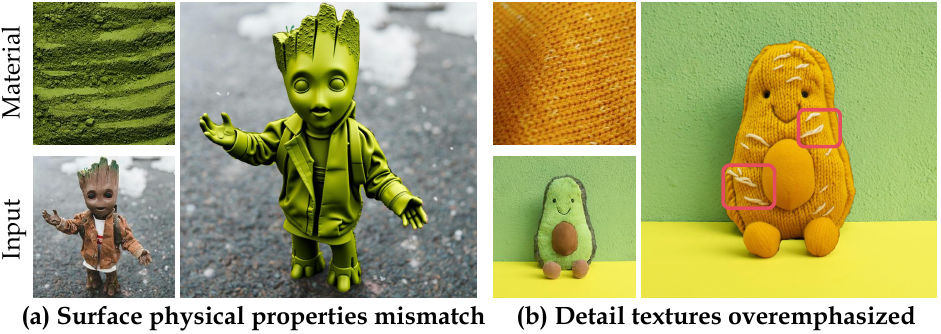}
\caption{Limitations. When encountering the extreme cases shown in the figure, DealMaTe produce failure and ambiguous material transfer results.
}
\label{fig:failure_cases}
\end{figure}

%% file: Figures/failure_condition.tex
\begin{figure}
\centering
\includegraphics[width=\linewidth]{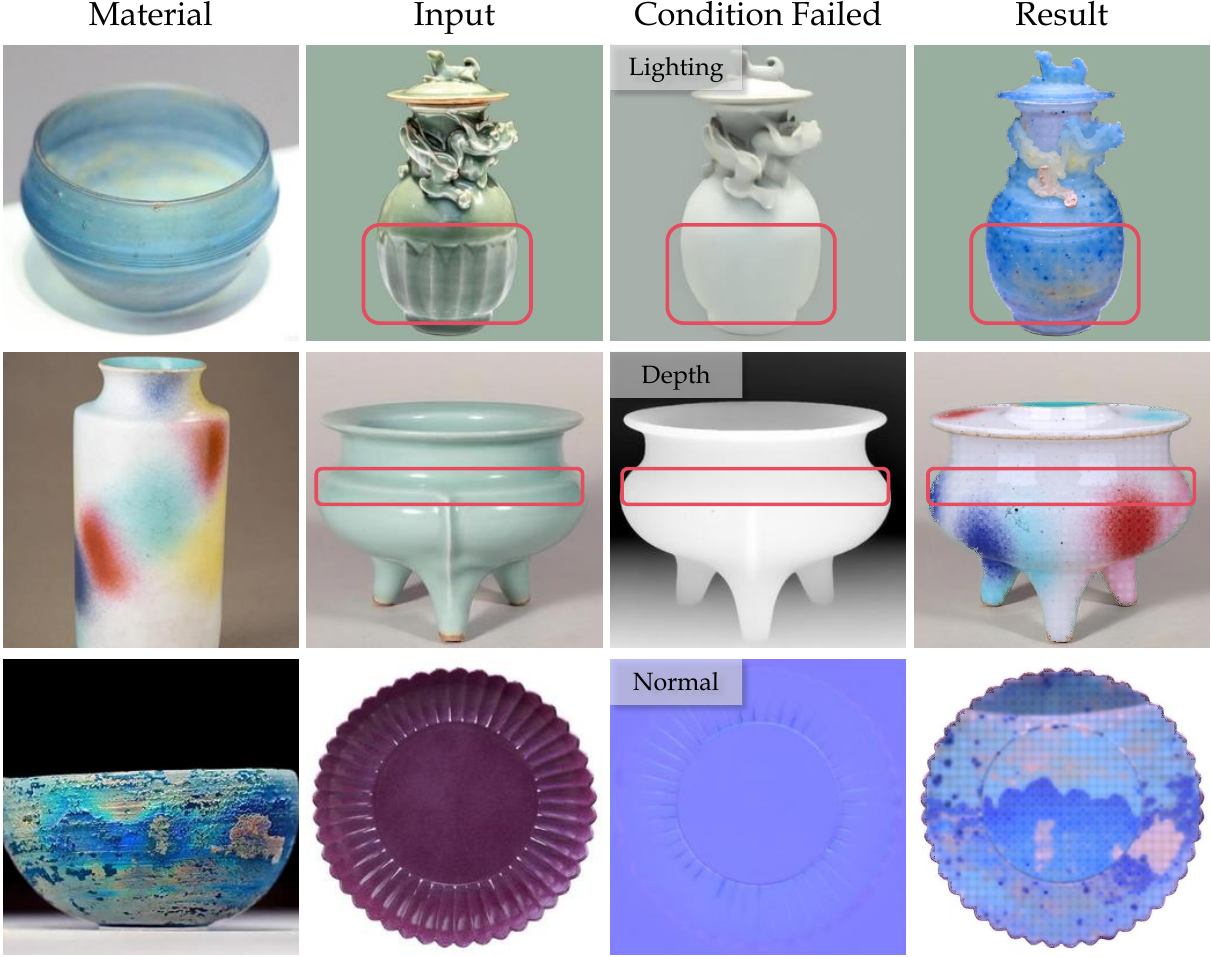}
\caption{Failure cases due to inaccurate geometric conditions estimation.}
\label{fig:failure_condition}
\end{figure}

%% file: Figures/discussion.tex
\begin{figure}[t]
\centering
\includegraphics[width= \linewidth]{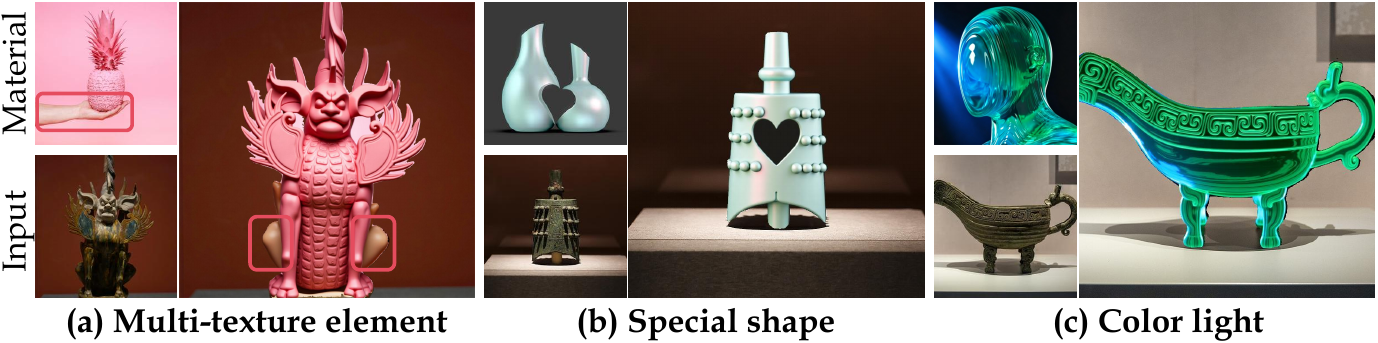}
\caption{Discussions. The results when the material diagram contains multiple types of materials or includes information beyond the materials.
}
\label{fig:discussion}
\end{figure}

%% file: Sections/6_conclusion.tex
\section{Conclusion}
Existing methods rely on image fine-tuning or complex architectures with reference networks, facing challenges such as text dependency, high computational costs, and insufficient structure preservation. 
To solve these issues, we propose the promotion framework DealMaTe. It establishes semantic alignment across material, depth, normal, and lighting images, allowing for material transfer from any 2D image to the target item.
The lightweight Multi-Dim 3D Shader LoRA module provides stable 3D information support without modifying the base model weights, ensuring harmonious and stable transfer results. Additionally, shader causal attention and KV caching optimize the attention mechanism, reducing inference latency and maintaining low architectural complexity. Extensive tests demonstrate that DealMaTe provides cutting-edge performance with a compact and efficient architecture.
In the future, we will further extend the 2D material transfer architecture to 3D objects and scenes.

%% file: sample-base.bib
@article{deschaintre2018single,
  title={Single-image svbrdf capture with a rendering-aware deep network},
  author={Deschaintre, Valentin and Aittala, Miika and Durand, Fredo and Drettakis, George and Bousseau, Adrien},
  journal={ACM Transactions on Graphics (TOG)},
  volume={37},
  number={4},
  pages={1--15},
  year={2018},
  publisher={ACM New York, NY, USA}
}

@inproceedings{rombach2022high,
  title={High-resolution image synthesis with latent diffusion models},
  author={Rombach, Robin and Blattmann, Andreas and Lorenz, Dominik and Esser, Patrick and Ommer, Bj{\"o}rn},
  booktitle={Proceedings of the IEEE/CVF Conference on Computer Vision and Pattern Recognition},
  pages={10684--10695},
  year={2022}
}

@inproceedings{esser2024scaling,
  title={Scaling Rectified Flow Transformers for High-Resolution Image Synthesis},
  author={Esser, Patrick and Kulal, Sumith and Blattmann, Andreas and Entezari, Rahim and M{\"u}ller, Jonas and Saini, Harry and Levi, Yam and Lorenz, Dominik and Sauer, Axel and Boesel, Frederic and others},
  booktitle={International Conference on Machine Learning},
  pages={12606--12633},
  year={2024},
  organization={PMLR}
}

@inproceedings{peebles2023scalable,
  title={Scalable diffusion models with transformers},
  author={Peebles, William and Xie, Saining},
  booktitle={Proceedings of the IEEE/CVF International Conference on Computer Vision},
  pages={4195--4205},
  year={2023}
}

@inproceedings{chen2023pixart,
  title={Pixart-$\alpha$: Fast training of diffusion transformer for photorealistic text-to-image synthesis},
  author={Chen, Junsong and Jincheng, YU and Chongjian, GE and Yao, Lewei and Xie, Enze and Wang, Zhongdao and Kwok, James and Luo, Ping and Lu, Huchuan and Li, Zhenguo},
  booktitle={International Conference on Learning Representations},
  year={2024}
}

@misc{flux2024,
    author={Black Forest Labs},
    title={FLUX},
    howpublished={\url{https://github.com/black-forest-labs/flux}},
    year={2024}
}

@inproceedings{ruiz2023dreambooth,
  title={Dreambooth: Fine tuning text-to-image diffusion models for subject-driven generation},
  author={Ruiz, Nataniel and Li, Yuanzhen and Jampani, Varun and Pritch, Yael and Rubinstein, Michael and Aberman, Kfir},
  booktitle={Proceedings of the IEEE/CVF Conference on Computer Vision and Pattern Recognition},
  pages={22500--22510},
  year={2023}
}

@inproceedings{galimage,
  title={An Image is Worth One Word: Personalizing Text-to-Image Generation using Textual Inversion},
  author={Gal, Rinon and Alaluf, Yuval and Atzmon, Yuval and Patashnik, Or and Bermano, Amit Haim and Chechik, Gal and Cohen-or, Daniel},
  booktitle={International Conference on Learning Representations},
  year={2023}
}

@inproceedings{hertz2022prompt,
  title={Prompt-to-Prompt Image Editing with Cross-Attention Control},
  author={Hertz, Amir and Mokady, Ron and Tenenbaum, Jay and Aberman, Kfir and Pritch, Yael and Cohen-or, Daniel},
  booktitle={International Conference on Learning Representations},
  year={2023}
}

@inproceedings{mou2024t2i,
  title={T2i-adapter: Learning adapters to dig out more controllable ability for text-to-image diffusion models},
  author={Mou, Chong and Wang, Xintao and Xie, Liangbin and Wu, Yanze and Zhang, Jian and Qi, Zhongang and Shan, Ying},
  booktitle={Proceedings of the AAAI Conference on Artificial Intelligence},
  volume={38},
  number={5},
  pages={4296--4304},
  year={2024}
}

@article{ye2023ip-adapter,
  title={IP-Adapter: Text Compatible Image Prompt Adapter for Text-to-Image Diffusion Models},
  author={Ye, Hu and Zhang, Jun and Liu, Sibo and Han, Xiao and Yang, Wei},
  journal={arXiv preprint arxiv:2308.06721},
  year={2023}
}

@inproceedings{sharma2024alchemist,
  title={Alchemist: Parametric control of material properties with diffusion models},
  author={Sharma, Prafull and Jampani, Varun and Li, Yuanzhen and Jia, Xuhui and Lagun, Dmitry and Durand, Fredo and Freeman, Bill and Matthews, Mark},
  booktitle={Proceedings of the IEEE/CVF Conference on Computer Vision and Pattern Recognition},
  pages={24130--24141},
  year={2024}
}

@inproceedings{chen2023text2tex,
  title={Text2tex: Text-driven texture synthesis via diffusion models},
  author={Chen, Dave Zhenyu and Siddiqui, Yawar and Lee, Hsin-Ying and Tulyakov, Sergey and Nie{\ss}ner, Matthias},
  booktitle={Proceedings of the IEEE/CVF International Conference on Computer Vision},
  pages={18558--18568},
  year={2023}
}

@article{garifullin2025materialfusion,
  title={MaterialFusion: High-Quality, Zero-Shot, and Controllable Material Transfer with Diffusion Models},
  author={Kamil Garifullin and Maxim Nikolaev and Andrey Kuznetsov and Aibek Alanov},
  journal={arXiv preprint arXiv:2502.06606},
  year={2025}
}

@inproceedings{richardson2023texture,
  title={Texture: Text-guided texturing of 3d shapes},
  author={Richardson, Elad and Metzer, Gal and Alaluf, Yuval and Giryes, Raja and Cohen-Or, Daniel},
  booktitle={ACM SIGGRAPH Conference Proceedings},
  pages={1--11},
  year={2023}
}

@inproceedings{yeh2024texturedreamer,
  title={Texturedreamer: Image-guided texture synthesis through geometry-aware diffusion},
  author={Yeh, Yu-Ying and Huang, Jia-Bin and Kim, Changil and Xiao, Lei and Nguyen-Phuoc, Thu and Khan, Numair and Zhang, Cheng and Chandraker, Manmohan and Marshall, Carl S and Dong, Zhao and others},
  booktitle={Proceedings of the IEEE/CVF Conference on Computer Vision and Pattern Recognition},
  pages={4304--4314},
  year={2024}
}

@article{zhang2023prospect,
  title={Prospect: Prompt spectrum for attribute-aware personalization of diffusion models},
  author={Zhang, Yuxin and Dong, Weiming and Tang, Fan and Huang, Nisha and Huang, Haibin and Ma, Chongyang and Lee, Tong-Yee and Deussen, Oliver and Xu, Changsheng},
  journal={ACM Transactions on Graphics (TOG)},
  volume={42},
  number={6},
  pages={1--14},
  year={2023},
  publisher={ACM New York, NY, USA}
}

@inproceedings{wu2024u,
  title={U-VAP: User-specified Visual Appearance Personalization via Decoupled Self Augmentation},
  author={Wu, You and Liu, Kean and Mi, Xiaoyue and Tang, Fan and Cao, Juan and Li, Jintao},
  booktitle={Proceedings of the IEEE/CVF Conference on Computer Vision and Pattern Recognition},
  pages={9482--9491},
  year={2024}
}

@inproceedings{cheng2024zest,
  title={Zest: Zero-shot material transfer from a single image},
  author={Cheng, Ta-Ying and Sharma, Prafull and Markham, Andrew and Trigoni, Niki and Jampani, Varun},
  booktitle={European Conference on Computer Vision},
  pages={370--386},
  year={2024},
  organization={Springer}
}

@inproceedings{lipmanflow,
  title={Flow Matching for Generative Modeling},
  author={Lipman, Yaron and Chen, Ricky TQ and Ben-Hamu, Heli and Nickel, Maximilian and Le, Matthew},
  booktitle={International Conference on Learning Representations},
  year={2023}
}

@inproceedings{liuflow,
  title={Flow Straight and Fast: Learning to Generate and Transfer Data with Rectified Flow},
  author={Liu, Xingchao and Gong, Chengyue and others},
  booktitle={International Conference on Learning Representations},
  year={2023}
}

@inproceedings{pan2020multi,
  title={Multi-Modal Attention for Speech Emotion Recognition},
  author={Pan, Zexu and Luo, Zhaojie and Yang, Jichen and Li, Haizhou},
  booktitle={Proceedings of the Annual Conference of the International Speech Communication Association (INTERSPEECH)},
  pages={364--368},
  year={2020}
}

@article{su2024roformer,
  title={Roformer: Enhanced transformer with rotary position embedding},
  author={Su, Jianlin and Ahmed, Murtadha and Lu, Yu and Pan, Shengfeng and Bo, Wen and Liu, Yunfeng},
  journal={Neurocomputing},
  volume={568},
  pages={127063},
  year={2024},
  publisher={Elsevier}
}

@article{wang2004image,
  title={Image quality assessment: from error visibility to structural similarity},
  author={Wang, Zhou and Bovik, Alan C and Sheikh, Hamid R and Simoncelli, Eero P},
  journal={IEEE Transactions on Image Processing},
  volume={13},
  number={4},
  pages={600--612},
  year={2004},
  publisher={IEEE}
}

@inproceedings{zhang2018unreasonable,
  title={The unreasonable effectiveness of deep features as a perceptual metric},
  author={Zhang, Richard and Isola, Phillip and Efros, Alexei A and Shechtman, Eli and Wang, Oliver},
  booktitle={Proceedings of the IEEE/CVF Conference on Computer Vision and Pattern Recognition},
  pages={586--595},
  year={2018}
}

@inproceedings{radford2021learning,
  title={Learning transferable visual models from natural language supervision},
  author={Radford, Alec and Kim, Jong Wook and Hallacy, Chris and Ramesh, Aditya and Goh, Gabriel and Agarwal, Sandhini and Sastry, Girish and Askell, Amanda and Mishkin, Pamela and Clark, Jack and others},
  booktitle={International Conference on Machine Learning},
  pages={8748--8763},
  year={2021},
  organization={PMLR}
}

@misc{unsplash,
  author={Unsplash Inc.},
  title={Unsplash},
  howpublished={\url{https://unsplash.com/}},
  year={2025}
}

@misc{control_v11f1p_sd15_depth,
  author={Lvmin Zhang},
  title={Controlnet-v1.1-depth},
  howpublished={\url{https://huggingface.co/lllyasviel/control_v11f1p_sd15_depth}},
  year={2023}
}

@misc{controlnet-depth-sdxl-1.0,
  author={Diffusers},
  title={controlnet-depth-sdxl-1.0},
  howpublished={\url{https://huggingface.co/diffusers/controlnet-depth-sdxl-1.0}},
  year={2023}
}

@article{tony,
  title={WYSIWYG Design of Hypnotic Line Art},
  author={Yeh, Chih-Kuo and Liu, Zhanping and Lin, I-Hsuan and Zhang, Eugene and Lee, Tong-Yee},
  journal={IEEE Transactions on Visualization and Computer Graphics},
  volume={28},
  number={6},
  pages={2517--2529},
  year={2020},
  publisher={IEEE}
}

@article{hu2019novel,
  title={A novel framework for inverse procedural texture modeling},
  author={Hu, Yiwei and Dorsey, Julie and Rushmeier, Holly},
  journal={ACM Transactions on Graphics (TOG)},
  volume={38},
  number={6},
  pages={1--14},
  year={2019},
  publisher={ACM New York, NY, USA}
}

@article{hu2022inverse,
  title={An inverse procedural modeling pipeline for svbrdf maps},
  author={Hu, Yiwei and He, Chengan and Deschaintre, Valentin and Dorsey, Julie and Rushmeier, Holly},
  journal={ACM Transactions on Graphics (TOG)},
  volume={41},
  number={2},
  pages={1--17},
  year={2022},
  publisher={ACM New York, NY}
}

@article{henzler2021generative,
  title={Generative modelling of BRDF textures from flash images},
  author={Henzler, Philipp and Deschaintre, Valentin and Mitra, Niloy J and Ritschel, Tobias},
  journal={ACM Transactions on Graphics (TOG)},
  volume={40},
  number={6},
  pages={1--13},
  year={2021},
  publisher={ACM New York, NY, USA}
}

@inproceedings{songdenoising,
  title={Denoising Diffusion Implicit Models},
  author={Song, Jiaming and Meng, Chenlin and Ermon, Stefano},
  booktitle={International Conference on Learning Representations},
  year={2021}
}

@article{huang2024diffstyler,
  title={Diffstyler: Controllable dual diffusion for text-driven image stylization},
  author={Huang, Nisha and Zhang, Yuxin and Tang, Fan and Ma, Chongyang and Huang, Haibin and Dong, Weiming and Xu, Changsheng},
  journal={IEEE Transactions on Neural Networks and Learning Systems},
  year={2024},
  publisher={IEEE}
}

@article{huang2025creativesynth,
  title={CreativeSynth: Cross-Art-Attention for Artistic Image Synthesis With Multimodal Diffusion},
  author={Huang, Nisha and Dong, Weiming and Zhang, Yuxin and Tang, Fan and Li, Ronghui and Ma, Chongyang and Li, Xiu and Lee, Tong-Yee and Xu, Changsheng},
  journal={IEEE Transactions on Visualization and Computer Graphics},
  year={2025},
  publisher={IEEE}
}

@inproceedings{cheng2025marble,
  title={MARBLE: Material Recomposition and Blending in CLIP-Space},
  author={Cheng, Ta Ying and Sharma, Prafull and Boss, Mark and Jampani, Varun},
  booktitle={Proceedings of the IEEE/CVF Conference on Computer Vision and Pattern Recognition},
  pages={13061--13071},
  year={2025}
}

@article{comanici2025gemini,
  title={Gemini 2.5: Pushing the frontier with advanced reasoning, multimodality, long context, and next generation agentic capabilities},
  author={Comanici, Gheorghe and Bieber, Eric and Schaekermann, Mike and Pasupat, Ice and Sachdeva, Noveen and Dhillon, Inderjit and Blistein, Marcel and Ram, Ori and Zhang, Dan and Rosen, Evan and others},
  journal={arXiv preprint arXiv:2507.06261},
  year={2025}
}

@inproceedings{brooks2023instructpix2pix,
  title={Instructpix2pix: Learning to follow image editing instructions},
  author={Brooks, Tim and Holynski, Aleksander and Efros, Alexei A},
  booktitle={Proceedings of the IEEE/CVF Conference on Computer Vision and Pattern Recognition},
  pages={18392--18402},
  year={2023}
}

@InProceedings{Huang_2025_ICCV,
    author    = {Huang, Nisha and Liu, Henglin and Lin, Yizhou and Huang, Kaer and Chen, Chubin and Guo, Jie and Lee, Tong-yee and Li, Xiu},
    title     = {MaTe: Images Are All You Need for Material Transfer via Diffusion Transformer},
    booktitle = {Proceedings of the IEEE/CVF International Conference on Computer Vision},
    month     = {October},
    year      = {2025},
    pages     = {15117-15126}
}

@article{Guarnera16,
author = {Guarnera, D. and Guarnera, G.C. and Ghosh, A. and Denk, C. and Glencross, M.},
title = {BRDF Representation and Acquisition},
journal = {Computer Graphics Forum},
volume = {35},
number = {2},
pages = {625-650},
doi = {https://doi.org/10.1111/cgf.12867},
year = {2016}
}

@inproceedings{asselin2020deep,
  title={Deep SVBRDF estimation on real materials},
  author={Asselin, Louis-Philippe and Laurendeau, Denis and Lalonde, Jean-Francois},
  booktitle={International Conference on 3D Vision (3DV)},
  pages={1157--1166},
  year={2020},
  organization={IEEE}
}

@inproceedings{deschaintre2021deep,
  title={Deep polarization imaging for 3d shape and svbrdf acquisition},
  author={Deschaintre, Valentin and Lin, Yiming and Ghosh, Abhijeet},
  booktitle={Proceedings of the IEEE/CVF Conference on Computer Vision and Pattern Recognition},
  pages={15567--15576},
  year={2021}
}

@article{gao2019deep,
  title={Deep inverse rendering for high-resolution SVBRDF estimation from an arbitrary number of images.},
  author={Gao, Duan and Li, Xiao and Dong, Yue and Peers, Pieter and Xu, Kun and Tong, Xin},
  journal={ACM Transactions on Graphics (TOG)},
  volume={38},
  number={4},
  pages={134--1},
  year={2019}
}

@inproceedings{martin2022materia,
  title={MaterIA: Single Image High-Resolution Material Capture in the Wild},
  author={Martin, Rosalie and Roullier, Arthur and Rouffet, Romain and Kaiser, Adrien and Boubekeur, Tamy},
  booktitle={Computer Graphics Forum},
  volume={41},
  number={2},
  pages={163--177},
  year={2022},
  organization={Wiley Online Library}
}

@article{li2017modeling,
  title={Modeling surface appearance from a single photograph using self-augmented convolutional neural networks},
  author={Li, Xiao and Dong, Yue and Peers, Pieter and Tong, Xin},
  journal={ACM Transactions on Graphics (TOG)},
  volume={36},
  number={4},
  pages={1--11},
  year={2017},
  publisher={ACM New York, NY, USA}
}

@inproceedings{vecchio2021surfacenet,
  title={Surfacenet: Adversarial svbrdf estimation from a single image},
  author={Vecchio, Giuseppe and Palazzo, Simone and Spampinato, Concetto},
  booktitle={Proceedings of the IEEE/CVF International Conference on Computer Vision},
  pages={12840--12848},
  year={2021}
}

@inproceedings{deschaintre2020guided,
  title={Guided fine-tuning for large-scale material transfer},
  author={Deschaintre, Valentin and Drettakis, George and Bousseau, Adrien},
  booktitle={Computer Graphics Forum},
  volume={39},
  number={4},
  pages={91--105},
  year={2020},
  organization={Wiley Online Library}
}

@inproceedings{rodriguez2023umat,
  title={Umat: Uncertainty-aware single image high resolution material capture},
  author={Rodriguez-Pardo, Carlos and Dominguez-Elvira, Henar and Pascual-Hernandez, David and Garces, Elena},
  booktitle={Proceedings of the IEEE/CVF Conference on Computer Vision and Pattern Recognition},
  pages={5764--5774},
  year={2023}
}

@inproceedings{lopes2024material,
  title={Material palette: Extraction of materials from a single image},
  author={Lopes, Ivan and Pizzati, Fabio and de Charette, Raoul},
  booktitle={Proceedings of the IEEE/CVF Conference on Computer Vision and Pattern Recognition},
  pages={4379--4388},
  year={2024}
}

@article{ma2025materialpicker,
  title={MaterialPicker: Multi-Modal DiT-Based Material Generation},
  author={Ma, Xiaohe and Deschaintre, Valentin and Ha{\v{s}}an, Milo{\v{s}} and Luan, Fujun and Zhou, Kun and Wu, Hongzhi and Hu, Yiwei},
  journal={ACM Transactions on Graphics (TOG)},
  volume={44},
  number={4},
  pages={1--12},
  year={2025},
  publisher={ACM New York, NY, USA}
}

@article{rosenberger2009layered,
  title={Layered shape synthesis: automatic generation of control maps for non-stationary textures},
  author={Rosenberger, Amir and Cohen-Or, Daniel and Lischinski, Dani},
  journal={ACM Transactions on Graphics (TOG)},
  volume={28},
  number={5},
  pages={1--9},
  year={2009},
  publisher={ACM New York, NY, USA}
}

@inproceedings{li2022scraping,
  title={Scraping textures from natural images for synthesis and editing},
  author={Li, Xueting and Wang, Xiaolong and Yang, Ming-Hsuan and Efros, Alexei A and Liu, Sifei},
  booktitle={European Conference on Computer Vision},
  pages={391--408},
  year={2022},
  organization={Springer}
}

@inproceedings{cazenavette2022wearable,
  title={Wearable ImageNet: Synthesizing tileable textures via dataset distillation},
  author={Cazenavette, George and Wang, Tongzhou and Torralba, Antonio and Efros, Alexei A and Zhu, Jun-Yan},
  booktitle={Proceedings of the IEEE/CVF Conference on Computer Vision and Pattern Recognition},
  pages={2278--2282},
  year={2022}
}

@InProceedings{ke2023repurposing,
  title={Repurposing Diffusion-Based Image Generators for Monocular Depth Estimation},
  author={Bingxin Ke and Anton Obukhov and Shengyu Huang and Nando Metzger and Rodrigo Caye Daudt and Konrad Schindler},
  booktitle = {Proceedings of the IEEE/CVF Conference on Computer Vision and Pattern Recognition},
  year={2024}
}

@misc{ke2025marigold,
  title={Marigold: Affordable Adaptation of Diffusion-Based Image Generators for Image Analysis},
  author={Bingxin Ke and Kevin Qu and Tianfu Wang and Nando Metzger and Shengyu Huang and Bo Li and Anton Obukhov and Konrad Schindler},
  year={2025},
  eprint={2505.09358},
  archivePrefix={arXiv},
  primaryClass={cs.CV}
}

@article{texsynth,
  title={Synthesis of complex image appearance from limited exemplars},
  author={Diamanti, Olga and Barnes, Connelly and Paris, Sylvain and Shechtman, Eli and Sorkine-Hornung, Olga},
  journal={ACM Transactions on Graphics (TOG)},
  volume={34},
  number={2},
  pages={1--14},
  year={2015},
  publisher={ACM New York, NY, USA}
}

@inproceedings{tan2025ominicontrol,
  title={Ominicontrol: Minimal and universal control for diffusion transformer},
  author={Tan, Zhenxiong and Liu, Songhua and Yang, Xingyi and Xue, Qiaochu and Wang, Xinchao},
  booktitle={Proceedings of the IEEE/CVF International Conference on Computer Vision},
  pages={14940--14950},
  year={2025}
}

@inproceedings{hulora,
  title={LoRA: Low-Rank Adaptation of Large Language Models},
  author={Hu, Edward J and Wallis, Phillip and Allen-Zhu, Zeyuan and Li, Yuanzhi and Wang, Shean and Wang, Lu and Chen, Weizhu and others},
  booktitle={International Conference on Learning Representations},
  year={2022}
}

@inproceedings{wimbauer2024cache,
  title={Cache me if you can: Accelerating diffusion models through block caching},
  author={Wimbauer, Felix and Wu, Bichen and Schoenfeld, Edgar and Dai, Xiaoliang and Hou, Ji and He, Zijian and Sanakoyeu, Artsiom and Zhang, Peizhao and Tsai, Sam and Kohler, Jonas and others},
  booktitle={Proceedings of the IEEE/CVF Conference on Computer Vision and Pattern Recognition},
  pages={6211--6220},
  year={2024}
}

@article{ma2024learning,
  title={Learning-to-cache: Accelerating diffusion transformer via layer caching},
  author={Ma, Xinyin and Fang, Gongfan and Bi Mi, Michael and Wang, Xinchao},
  journal={Advances in Neural Information Processing Systems},
  volume={37},
  pages={133282--133304},
  year={2024}
}

@inproceedings{zhang2023adding,
  title={Adding conditional control to text-to-image diffusion models},
  author={Zhang, Lvmin and Rao, Anyi and Agrawala, Maneesh},
  booktitle={Proceedings of the IEEE/CVF International Conference on Computer Vision},
  pages={3836--3847},
  year={2023}
}

@inproceedings{psnr,
  title={Image quality metrics: PSNR vs. SSIM},
  author={Hore, Alain and Ziou, Djemel},
  booktitle={2010 20th international conference on pattern recognition},
  pages={2366--2369},
  year={2010},
  organization={IEEE}
}

@article{fu2023dreamsim,
  title={DreamSim: Learning New Dimensions of Human Visual Similarity using Synthetic Data},
  author={Fu, Stephanie and Tamir, Netanel and Sundaram, Shobhita and Chai, Lucy and Zhang, Richard and Dekel, Tali and Isola, Phillip},
  journal={Advances in Neural Information Processing Systems},
  volume={36},
  pages={50742--50768},
  year={2023}
}
